\documentclass[printer]{aa}

\usepackage{amsmath}
\usepackage{graphicx}
\usepackage{appendix}
\usepackage{txfonts}
\usepackage{color}
\usepackage{natbib}

\bibpunct{(}{)}{;}{a}{}{,} 

\newcommand{\eps}{\varepsilon}
\newcommand{\freq}{\omega}

\newcommand{\beq}{\begin{equation}}
\newcommand{\eeq}{\end{equation}}
\newcommand{\dd}{{\rm d}}

\begin{document}

\title{Broadband spectroscopy of astrophysical ice analogs.\\ I. Direct measurement of complex refractive index of CO ice using terahertz
time-domain spectroscopy}

 \author{B. M. Giuliano\inst{\ref{MPI}}, A. A. Gavdush\inst{\ref{GPI},\ref{BMSTU}}, B. M\"uller\inst{\ref{MPI}},
 K. I. Zaytsev\inst{\ref{GPI},\ref{BMSTU}},T. Grassi\inst{\ref{USM},\ref{ECOSU}}, A. V. Ivlev\inst{\ref{MPI}},
 M. E. Palumbo\inst{\ref{OAC}}, G. A. Baratta\inst{\ref{OAC}},
 C. Scir\`e\inst{\ref{OAC}}, G. A. Komandin\inst{\ref{GPI}}, S. O. Yurchenko\inst{\ref{BMSTU}}, 
 P. Caselli\inst{\ref{MPI}}}

 \offprints{Barbara M. Giuliano, Alexei V. Ivlev}

 \institute{Max-Planck-Institut f\"ur extraterrestrische Physik,
 Gie{\ss}enbachstrasse 1, D-85748 Garching, Germany \label{MPI}
 \and Prokhorov General Physics Institute of the Russian Academy of Sciences, 119991 Moscow, Russia \label{GPI}
 \and Bauman Moscow State Technical University, 105005 Moscow, Russia \label{BMSTU}
 \and Universit\"ats-Sternwarte M\"unchen, Scheinerstr. 1, D-81679 M\"unchen, Germany \label{USM}
 \and Excellence Cluster Origin and Structure of the Universe, Boltzmannstr. 2, D-85748 Garching bei M\"unchen, Germany \label{ECOSU}
 \and INAF - Osservatorio Astrofisico di Catania, via Santa Sofia 78, I-95123, Catania, Italy\label{OAC} }

\authorrunning{B. M. Giuliano et al.}
\titlerunning{THz complex refractive index of CO ice}

\abstract
{Reliable, directly measured optical properties of astrophysical ice analogs in the infrared (IR) and terahertz
(THz) range are missing. These parameters are of great importance to model the dust continuum radiative transfer 
in dense and cold regions, where thick ice mantles are present, and are necessary for the interpretation of future 
observations planned in the far-IR region.}
{Coherent THz radiation allows direct measurement of the complex dielectric function (refractive index) of
astrophysically relevant ice species in the THz range.}
{The time-domain waveforms and the frequency-domain spectra of reference samples of CO ice, deposited at a temperature of
28.5~K and annealed to 33~K at different thicknesses, have been recorded. A new algorithm is developed to reconstruct the
real and imaginary parts of the refractive index from the time-domain THz data.}
{The complex refractive index in the wavelength range of 1~mm--150~$\mu$m (0.3--2.0~THz) has been determined for the
studied ice samples, and compared with available data found in the literature.}
{The developed algorithm of reconstructing the real and imaginary parts of the refractive index 
from the time-domain THz data enables, for the first time, the determination of optical properties of astrophysical ice analogs 
without using the Kramers-Kronig relations. The obtained data provide a benchmark to interpret the observational data from current 
ground based facilities as well as future space telescope missions, 
and have been used to estimate the opacities of the dust grains in presence of CO ice mantles.}


\maketitle

%
%
\section{Introduction}
\label{intro}

One of the main problems in unraveling the chemical and physical properties of molecular clouds, in which the star and planet formation process takes place, is to estimate correctly the amount of gas contained.
The difficulties in the direct observation of molecular hydrogen constrain the possibility to calculate the total mass of a cloud. An easy alternative could be to use carbon monoxide as a tracer of molecular gas, 
but in dense and cold regions of the interstellar medium and protoplanetary disks, CO is not a good tracer of gas mass because CO molecules preferentially resides onto dust grains, 
forming thick icy mantles \citep[e.g ][]{1998A&A...338L..63D,1999ApJ...523L.165C}.

In alternative, the dust continuum emission is the best available tool to compute a molecular cloud mass, if dust opacities are known.

The advent of ALMA and NOEMA facilities offer the possibility to observe the dust continuum emission in the millimeter and sub-millimeter part of the electromagnetic spectrum, with very high angular resolution 
and sensitivity. However, to properly model the dust continuum emission it is necessary to have information about its grain size distribution as well as its chemical composition, since the dust opacity depends directly on these paramenters. 
If we take into account that the dust grains can be covered by ice mantles, at the center of prestellar cores or in protoplanetary disk midplanes, we also need to investigate how the presence of ices 
is changing the dust opacities.

Unfortunately, no experimental data are available for these cases, and the interpretation of the dust continuum emission measurements relais on calculated opacity values, as the ones tablulated in \citet{1994A&A...291..943O}.
The goal of our study is to provide laboratory data on the optical properties of CO ice and utilize these data to calculate the opacities of dust grains covered by CO ice mantles. 
We will compare the opacity values obtained by our study to those available in the literature.

The presently available set of data focus mainly on the determination of the optical constants in the visible and mid-IR range
\citep{AstroPhysJSupplSer.86.2.713.1993,1997A&A...328..649E,CO_Baratta_1998,2005A&A...435..587L,2006A&A...445..959D,2006PCCP....8..279P,2008JGRD..11314220W,2009ApJ...701.1347M}.

FIR studies on spectral properties of molecular solids, without deriving optical constants, started early, to deepen the understanding of the infrared-active lattice vibrations of simple species. 
\citet{1966JChPh..45.4359A} provided studies on frequencies of CO and N$_2$  in the range 40-100 cm$^{-1}$, and \citet{1967JChPh..46.3991R} complemented the available information with CO, N$_2$ and CO$_2$ intensity studies in the same frequency range.
Moore and Hudson published in 1994 a comprehensive study of FIR spectra of cosmic type ices, including ice mixtures. 
These data include the analysis of amorphous and crystalline phases of the pure molecular ices, and the authors discuss the implications of the results on the identification based on astronomical observation. 
An estimation of the band strengths in the FIR region for pure ices and ice mixtures relevant for astrophysical environments can be found in \citeauthor{2014A&A...565A.108G} (\citeyear{2014A&A...565A.108G,2016A&A...592A..81G}).

Recently, the investigation of the THz spectroscopic properties of ice mantles
analogs has gained a considerable interest \citep{2014PCCP...16.3442A,C3FD00154G, PhysChemChemPhys.18.30.20199.2016}.
This technique allows to measure directly the intermolecular vibrations in the ice samples, related to the lattice structure, which can be connected to their large-scale structural changes and finally to their thermal history. 
On the contrary, the spectroscopic features measured in the MIR frequency range are indicative of the intramolecular vibrations of the sample, which can provide a wealth of information on the molecular identification and chemical reactivity. 
A comparison of our THz experimental data with that observed in the MIR range, could help us to reveal intra and inter molecular vibrations. However, this study is beyond the scope of this paper and will be addressed in our future investigations.

Nowadays, numerous spectroscopic methods are extensively used for dielectric measurements at THz frequencies
\citep{YunShikLeeBook2009}; particularly, we would mention the following ones: Fourier Transform Infrared (FTIR)
spectroscopy \citep{GriffithsFTIR_Book1986}, Backward-Wave Oscillator (BWO) spectroscopy \citep{IEEE_BWO_2013}, spectroscopy
based on photomixing \citep{CW_THz_REV_2011} or parametric conversion \citep{PC_Kawase_1996,CW_THz_OPO_2013}, and, finally,
THz time-domain spectroscopy (THz-TDS) \citep{Auston_TDS_1975, Grischkowsky_APL_1989}. These methods exploit either
continuous-wave or broadband sources, operate in different spectral ranges, and are characterized with different sensitivity
and performance. Among them, the THz-TDS seems to be the most appropriate for studying laboratory analogs of circumstellar
and interstellar ices. In contrast to other approaches, the THz-TDS yields detection of both amplitude and phase of
sub-picosecond THz pulses in a wide spectral range as a result of a single measurement; thus, reconstruction of the
dielectric response of a sample might be performed without using the Kramers-Kronig relations
\citep{PhysRev.161.1.143-155.1967} and involving additional physical assumptions. Furthermore, the THz-TDS yields
analysis of separate wavelets forming the time-domain response of a sample; thus, it is a powerful method for the
characterization of multilayer samples. Thereby, we have selected the THz-TDS as a spectroscopic technique for our
experiments.

We aim at the extension of the laboratory data in the far-IR/THz region, and we will show how the employment of the THz-TDS
is able to provide the direct measurement of the real and imaginary part of the refractive index of the ice sample. The
experimental and theoretical methods employed will be explained in Section~\ref{expe}, the results obtained and how these
data are relevant for astrophysical application will be presented in sections Section~\ref{results} and Section~\ref{disc},
respectively, the conclusions will be  illustrated in Section~\ref{conc}.

\section{Experimental and theoretical methods}\label{expe}

For this series of experiments a dedicated setup has been designed and developed in the laboratories of the Center for
Astrochemical Studies (CAS) located at the Max Planck Institute for Extraterrestrial Physics in Garching (Germany). The
setup is composed of a closed-cycle He cryocooler coupled to a THz time-domain spectrometer. The cryocooler vacuum
chamber is small enough to be hosted in the sample compartment of the THz spectrometer, and it is mounted on a motor
controlled translational stage which ensures the tuning of the cryostat position with respect to the THz beam. The details
of the main components of the apparatus and the ice growing procedure are given in the following subsections.

\subsection{Cryogenic setup}
\label{cryo}

The high power cryocooler has been purchased from Advanced Research Systems. The model chosen is designed to handle high heat
loads ensuring a fast cooling. It is equipped with a special interface capable of reducing the vibration transmitted from
the cold head to the sample holder at the nm level. This requirement is important in case of spectroscopic measurements in 
the THz frequency region, where the induced vibration of the sample can cause the increase of the noise level of the recorded spectra.

The cryostat is placed in a 15 cm diameter vacuum chamber, equipped with four ports for optical access as
well as for the gas inlet. The optical arrangement is designed to work in the transmission configuration. The optical
windows chosen for the measurements at the desired frequency range are made of high-resistivity float-zone silicon
(HRFZ-Si), purchased by Tydex. This material features high refractive index of $n_\mathrm{Si}=3.415$, negligible dispersion
and impressive transparency in the desired frequency range. The same material has been chosen as a substrate for the ice
growing. In order to suppress the Fabry-Perot resonances in the THz spectra, caused by multiple reflection of the THz pulse
within the windows, the thickness of the three Si windows must be different from each other. For this purpose we have chosen
to use the 1 mm thick Si window as substrate for the ice deposition, and use two windows of 2 mm and 3 mm thickness as
optical access to the THz beam.

\begin{figure}[!t]
    \centering
    \includegraphics[width=9.cm]{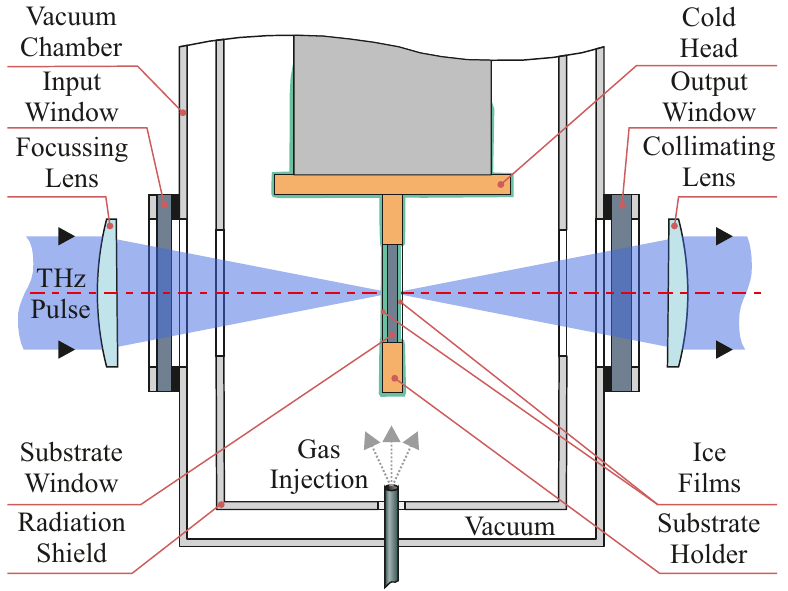}
    \caption{A sketch of the vacuum chamber of the cryostat coupled to the THz beam at the Center for Astrochemical Studies.}
    \label{chamber}
\end{figure}

We performed measurements with a slightly focused THz beam configuration in order to mitigate vignetting and diffraction of
the beam at the metal components of the vacuum chamber. A schematic overview of the chamber arrangement is sketched in
Fig. \ref{chamber}. The pumping station is composed of a turbomolecular pump (84 $\mathrm{ls}^{-1}$ nitrogen pumping
speed) combined with a backing rotary pump (5 $\mathrm{m}^{3}\mathrm{h}^{-1}$ pumping speed) providing a base pressure of
about $10^{-7}{\enskip}\mathrm{mbar}$. The minimum temperature measured at the sample holder in normal operation mode is 5
$\mathrm{K}$.

\subsection{THz time-domain spectrometer}
\label{THz}

The THz-TDS setup used for this work has been purchased from the company Batop GmbH. The model chosen
is the TDS-1008, with a customized sample compartment in order to allocate the cryostat. It is based on
two photoconductive antennas made of low-temperature-grown gallium arsenide (LT-GaAs), which constitute the emitter and the
detector of the THz pulse \citep{YunShikLeeBook2009}. The antennas are triggered by a femtosecond laser (TOPTICA, 95 fs, 780
nm) with a pulse repetition rate of 100 MHz and an average input power of 65 mW. Further details on the setup are provided in
the Appendix~\ref{App-A}.

Fig.~\ref{TDS}~(a) shows the optical path of the laser beam into the optical bench of the spectrometer. Panel~(b) in Fig.~\ref{TDS} shows standard
broadband Fourier spectra. The THz pulse registered with the beam path empty is then converted in the blue spectrum spanning
the frequency range of $0.05$ up to $3.5$~THz with maximal spectral amplitude centered at about $1.0$~THz. 
In TDS, the time-domain THz waveform is converted to the frequency domain using the direct Fourier transform, which yields the frequency-dependent amplitude and phase of the THz wavelet.
Since the frequency-domain data is calculated via the direct Fourier transform, the spectral resolution of measurements is determined as $\Delta \nu = 1 / \Delta T$,
where $\Delta T$ is a size of the of the time-domain apodization filter, chosen to avoid the edge effects (i.e. the Gibbs effect) in the frequency-domain. 
In our experiments we used the 35-ps Tukey apodization filter (see Appendix~\ref{App-B}), which yields the spectral resolution of about 0.03~THz.

The green spectrum has been recorded with the cryostat placed in the sample compartment. It is converted from a waveform which
contains both a first THz pulse (i.e. a ballistic one) and a train of satellite pulses originating from to the multiple THz
wave reflections within the windows. The ballistic THz pulse is delayed in the input/output windows and the substrate of the
vacuum chamber. The spectrum of this waveform is slightly suppressed due to the Fresnel losses and modulated due to the
interference of the ballistic pulse and the satellites.

In Fig.~\ref{TDS} (b), we show the shaded area at lower frequencies, where we expect growing distortions of the experimental data
caused by the THz beam diffraction on the $20$-mm-diameter aperture of the substrate. Assuming that the THz beam spot formed
at the substrate is diffraction-limited, the lateral intensity distribution in the spot is defined by the Bessel function of
the first kind \citep{WolfBorn.Book.1980}. The resulting width of the first intensity peak is approximately
$(3.8/{\pi})(f/D)(c/{\nu})$, where $D = 25$~mm and $f = 67$~mm stand for the diameter and the back focal distance of the
focusing lens, respectively. From this model, we deduce the critical frequency of $0.3$~THz, below which less than $95$\% of the beam
energy passes through the substrate aperture. Thereby, considering both the spectral sensitivity of our THz-TDS setup and the
diffraction limits, the spectral operation range of our experimental setup is approximately limited within $0.3$ to
$2.0$~THz.

The THz-TDS housing is kept under purging with cold nitrogen gas during the entire experiment, to mitigate the absorption
features due to the presence of atmospheric water in the THz beam path. The residual humidity measured at the sensor was
less then $10^{-3} \%$.

\begin{figure}[!t]
    \centering
    \includegraphics[width=9.cm]{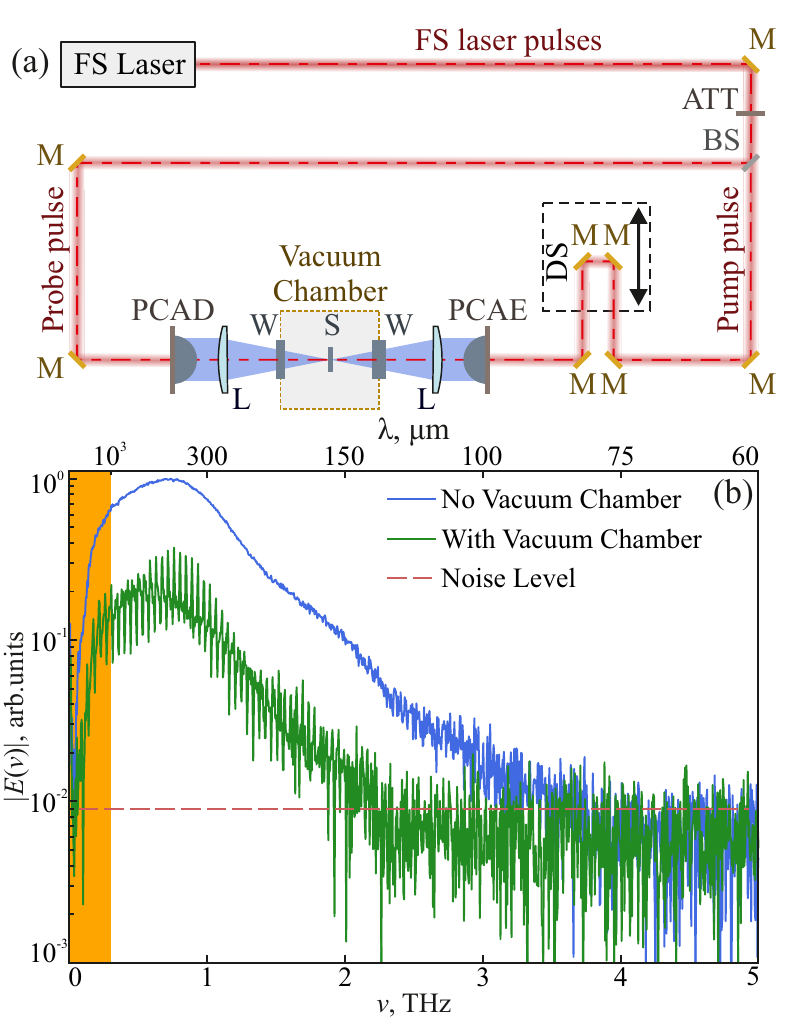}
    \caption{THz-TDS setup for the spectroscopy of ices. (a)~A schematic of the setup, where FS laser stands for the
    femtosecond laser, M stands for the optical mirrors, ATT stands for the attenuator of the laser beam intensity,
      BS stands for the optical beams splitter, DS stands for the mechanical double-pass delay stage, PCAE and PCAD stand
      for the photoconductive antenna-emitter and antenna-detector, respectively, L stand for the TPX lenses,
      S and W stand for the HRFZ-Si substrate and windows, respectively. (b)~Spectra of THz waveforms $E \left( \nu \right)$
      transmitted through the empty THz beam path or the THz beam path with the cryostat; the shaded region below
      $\approx0.3$~THz indicates the spectral range where distortions due to the THz beam diffraction on the aperture of
      the substrate are expected.}
    \label{TDS}
\end{figure}

\subsection{Ice preparation}
\label{ice}

The ices are prepared using standard technique in which the molecular sample in its gaseous form is allowed to enter the
vacuum chamber through a stainless steel 6 mm pipe. The gas flux is controlled by a metering valve. Once the gas is
expanding inside the vacuum chamber it condensates on the substrate.

For this set of experiments an ice thickness of the order of mm is required, to fulfill the sensitivity characteristic of
the THz-TDS setup. This value is orders of magnitude higher than the usual ice thickness reached using this deposition technique,
which is of the order of $\mu$m. To deposit such a thick ice in a reasonable amount of time, we have chosen fast deposition
conditions, in which a considerable amount of gas is introduced in the vacuum chamber. In these conditions it is very
difficult to obtain an ice sample which is homogeneous enough to determine its optical properties. To overcome this problem
the gas inlet characteristics must be set accordingly.

We decided to remove any directionality from the gas inlet, keeping the pipe end cut at the vacuum chamber wall, at a
distance of approximately 7 cm from the substrate (see Fig.~\ref{chamber}). This configuration creates an ice layer on each
side of the substrate. During the deposition, the pressure measured inside the chamber is approximately 10$^{-2}$~mbar. 
The ice deposition was divided in steps of 4, 5, and 6 min duration, up to a total of 30 min deposition time, in three different experiments. 
The final temperature was up to 28.5, 31.2 and 33.1 K for each step, at 4, 5, and 6 min deposition time, respectively. 
This increase is due to the condensation of the gas onto the cold surfaces of the cryostat, which is producing a heating rate too fast to be dynamically removed from the cooling system during the deposition. 
After each step, the THz spectrum has been recorded. This procedure was performed in order to rule out possible effects on the ice structure due to different deposition temperatures.
As interstellar ices can be commonly found at temperatures as low as 10 K, the temperature of the cold substrate has been kept at 14~K, 
which is the lowest temperature achievable in the setup in this configuration, due to the fact that the radiation shield of the sample holder must 
be removed to ensure that no directionality of the molecular beam is present. Before moving to the next deposition step, the system was allowed to 
thermalize and spectra recorded after each deposition step have been taken at a temperature of 14~K.

As stated in \citet{2016A&A...594A..80U} the analysis of the Raman and IR spectra of experiments performed at increasing tempertaures from 17 to 32~K show no profile 
variation in the band at 2140~cm$^{-1}$ which could be ascribed to a structural change in the ice morphology.

The waveform recorded in the time domain is compared, as reference for the measurements, with the
waveform recorded for the substrate without ice, kept at a reference temperature of 14~K as well. 
After the deposition was completed, we have measured the spectra in different regions of the sample, to ensure that
the  ice morphology is spatially homogeneous. The results obtained from the spectra measured on a grid of 11 points spaced
by 2~mm are in agreement within 10\%, indicating a uniform ice formation over the substrate.

\section{Derivation of the optical constants}\label{opt}

In order to determine the optical constants, the ice thickness must be known. The laser interference technique is a well-established method to estimate the thickness of an ice sample deposited on a substrate as a function of the time. 
The absolute accuracy of this method is approximately within 5\%, but the maximum CO ice thickness that we can measure with this technique is limited to 5 $\mu$m before the reflected laser signal becomes too weak to be detected, 
due to scattering losses occurring both in the bulk and on surface of the film. Thus, this technique is well suited for studying thin layers, when the ice thickness in total is below 10 $\mu$m (5 $\mu$m on each side of the substrate), 
but is not appropriate for experiments on thick ice samples. In turn, for the thickness estimation of the mm-size sample of ice featuring rather low THz absorption, 
we developed a model to perform an initial estimation of the optical properties and of the ice thickness directly from the recorded THz spectra, as described in the following subsections.

\subsection{Ice parameters modeling}
\label{model}

Our model aims to reconstruct the optical properties of ices, which are defined as follows
\begin{equation}
    n \left( \nu \right) = n' \left( \nu \right) - i n '' \left( \nu \right)
                         \equiv~n' \left( \nu \right) - i \frac{ c } {2 \pi \nu} \alpha \left( \nu \right),
    \label{EQ:ComplexRefractiveIndex}
\end{equation}
where $n'$ and $n''$ are the real and imaginary parts of the complex refractive index $n$, $c$ is the light speed, and
$\alpha$ is the amplitude absorption coefficient, which is defined as half of the value of the power absorption coefficient. 

Equivalently, we can write
\begin{equation}
    n^2 \left( \nu \right)\equiv \varepsilon \left( \nu \right)
                                   = \varepsilon' \left( \nu \right) - i \varepsilon '' \left( \nu \right),
    \label{EQ:ComplexDielectricPermittivity}
\end{equation}
where $\varepsilon '$ and $\varepsilon ''$ are the real and imaginary parts of the complex dielectric permittivity
$\varepsilon$.

The model is describing the THz wave propagation through the substrate with the ice deposited on both surfaces.
The reconstruction of the ice parameters proceeds following three main steps.

The first task is modeling the reference and sample waveforms. Because of the focused arrangement of the THz beams, the
electromagnetic wave is assumed to be planar and to interact with the sample interfaces at the normal angle of incidence.
This is a common and conventional assumption widely applied in dielectric spectroscopy
\citep{OptExp.15.7.4335.2007,JApplPhys.115.19.193105.2014}. It allows us to describe all the features of the THz pulse
interaction with the multilayer sample using the Fresnel formulas, which define the THz wave amplitude reflection at (and
transmission through) the interface between the media, and the Bouguer-Lambert-Beer law, which defines the absorption and the phase delay of the THz wave in a bulk medium. 
Further details on these assumptions are given in Appendix~\ref{App-B}.

\begin{figure}[!t]
    \centering
    \includegraphics[width=9.cm]{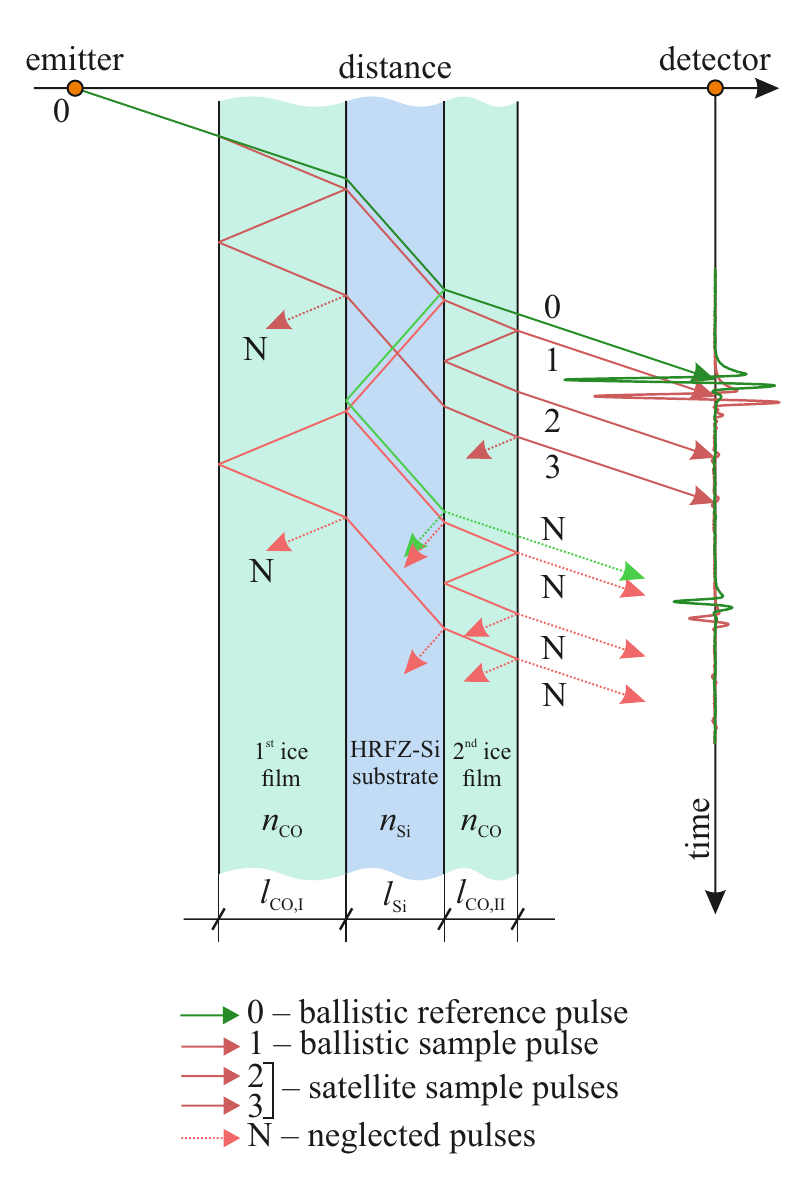}
    \caption{A time-distance diagram illustrating the THz wave propagation through the HRFZ-Si substrate with ice films
    deposited on its surfaces. Lines 0 to 3 illustrate the ballistic pulse and satellite pulses transmitted in the
    direction of the antenna-detector, N stands for unaccounted satellites with larger time delays. Solid lines represent the 
    pulses which have been used for the analysis, while dotted lines correspond to neglected pulses.}
    \label{layers}
\end{figure}

Figure~\ref{layers} represents the THz wave propagation through the three layers structure -- the first ice film, the
HRFZ-Si substrate, and the second ice film, where the symbols 0 to 3 and N correspond to different components of the plane
wave passing through the multilayered structure. As shown in Fig.~\ref{layers}, for the sample spectrum, we took into account
the contribution of the ballistic THz pulse (1) and the satellite pulses (2 and 3), caused by the multiple THz wave
reflection in the ice films. The mathematical description of the wave propagation can be found in Appendix~\ref{App-B}.

\begin{figure}[!t]
    \centering
    \includegraphics[width=9.cm]{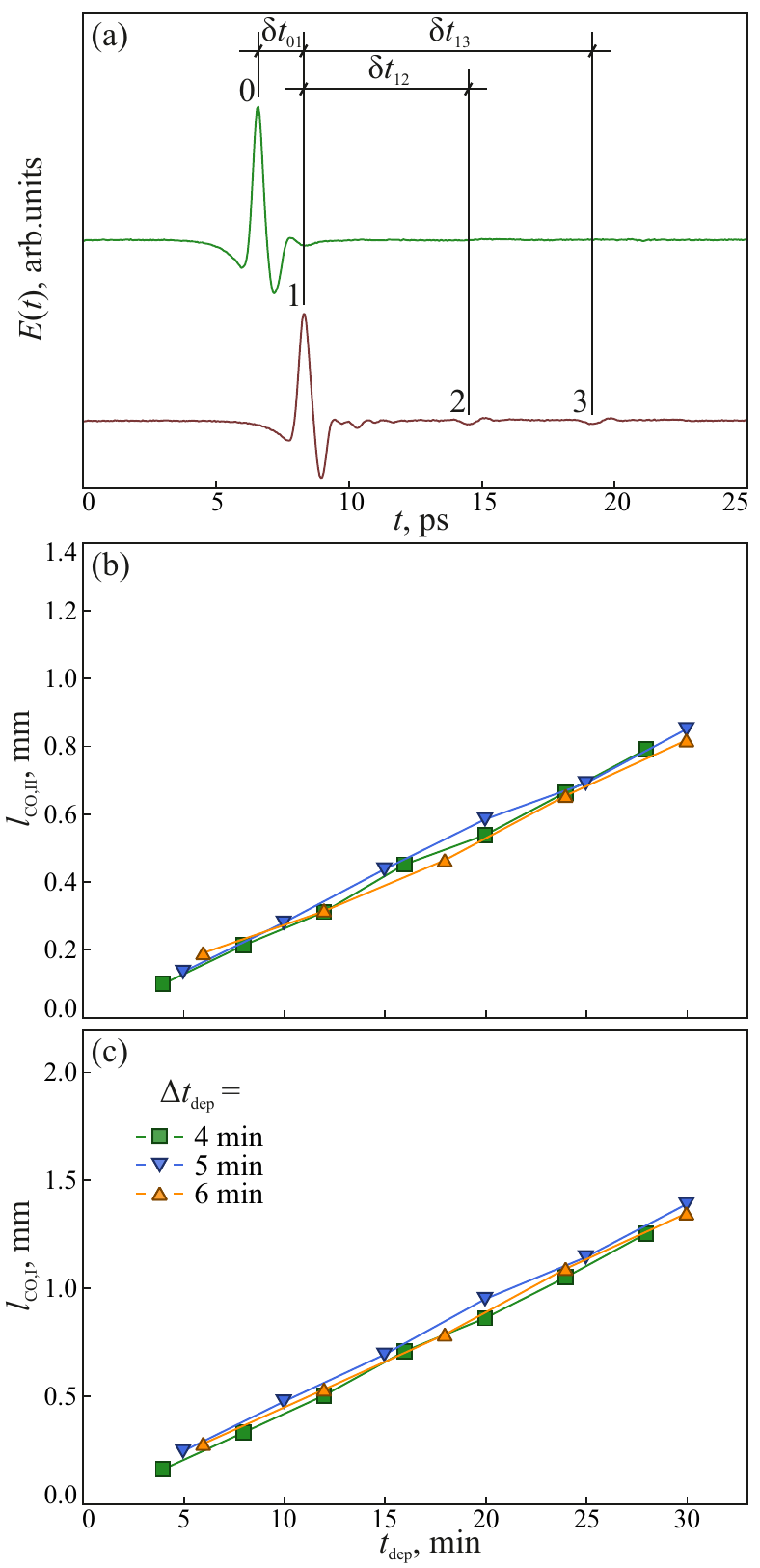}
    \caption{Calculation of the thicknesses $l_\mathrm{CO,I}$, $l_\mathrm{CO,II}$ and the complex refractive index
    $n$ of the CO ice films (see Fig.~\ref{layers}). (a)~Time delays between the ballistic THz pulses of the reference (0) and sample (1)
    waveforms, $\delta t_\mathrm{01}$; the first satellite pulse (2) and the ballistic pulse (1) of the sample waveform,
    $\delta t_\mathrm{12}$; and the second satellite pulse (3) and the ballistic pulse (1) of the sample waveform,
    $\delta t_\mathrm{13}$ (the pulses are marked as in Fig.~\ref{layers}). (b,c)~Estimates for the thicknesses of the
    two ice films as a function of the total deposition time $t_{\rm dep}$ for the different deposition
    intervals of $\Delta t_{\rm dep} = 4$, $5$ and $6$~min; the first assumption for the real part of the
    refractive index of ice is between $n'_\mathrm{init} = 1.230$ and $1.255 \pm 0.035$, for the imaginary part we first
    set $n''_\mathrm{init} = 0$.}
    \label{thick}
\end{figure}

The second step consists in estimating the initial thickness $l_\mathrm{CO,I}$, $l_\mathrm{CO,II}$ and the initial complex
refractive index $n_\mathrm{Init}$ of the two ice layers as shown in Fig.~\ref{thick}. The thickness estimation can be
derived from the time delay $\delta t_\mathrm{01}$, between the ballistic pulse of the reference and sample waveforms ($0$
and 1 in Fig.~\ref{thick}), the first satellite pulse (2) and the ballistic pulse (1) of the sample waveform $\delta
t_\mathrm{12}$, the second satellite pulse (3) and the ballistic pulse (1) of the sample waveform $\delta t_\mathrm{13}$, in
Fig.~\ref{thick}~(a). Since the HRFZ-Si has a very high refractive index, we consider that the refractive indexes of ice $n$
satisfies the inequality $n_\mathrm{0} = 1.0 < n < n_\mathrm{Si} = 3.415$, where $n_\mathrm{0}$ is the refractive index in vacuum.

Then, by neglecting imaginary parts in the complex refractive indexes of media, the first assumptions for the real part of 
the complex refractive index of ice films and both their thicknesses $l_\mathrm{CO,I}$ and $l_\mathrm{CO,II}$ are described as follows:
\begin{equation}
    n'_\mathrm{Init} = \frac{ \delta t_\mathrm{12} + \delta t_\mathrm{13} }{ \delta t_\mathrm{12}
    + \delta t_\mathrm{13} - 2\delta t_\mathrm{01} },
    \label{Eq:FirsAssumtionForTheRefractiveIndex}
\end{equation}

\begin{equation}
    l_\mathrm{CO,I} = \frac{c\: \delta t_\mathrm{13}}{ 2 n' },
    \qquad
    l_\mathrm{CO,II} = \frac{c\: \delta t_\mathrm{12}}{ 2 n' }.
    \label{Eq:FirsAssumtionForTheThicknesses}
\end{equation}

Eq.~\eqref{Eq:FirsAssumtionForTheRefractiveIndex} is obtained from a mathematical model of sample and reference waveform.
Futher details on the derivation can be found in Appendix~\ref{App-B}.
From Eq.~\eqref{Eq:FirsAssumtionForTheThicknesses} it is possible to obtain information only on the thicknesses of ice
films, while the identification of the specific layer (I or II) is not allowed. We observe a linear increase of the ice
thickness with the total deposition time. A first assumption for the real part of the complex refractive index of CO
ice is in the range of $n'_\mathrm{init} = 1.230$ to $1.255 \pm 0.035$ for all the considered deposition intervals; here,
the error accounts for an accuracy of the THz pulse peak position estimation. The first assumption for the imaginary part of
the complex refractive index of ice has been done considering $\alpha_\mathrm{init} = 0$; thus, $n''_\mathrm{init} = 0$.

Finally, from the first estimation of the ice thickness and complex refractive index, it is possible to reconstruct the THz
dielectric response of ice. The reconstruction procedure is reported in Appendix~\ref{App-B}, while the results are
summarized in Fig.~\ref{thick}. Panels~(b) and~(c) show the growth of the two CO ice layers in time ($t$), considering
different deposition steps of $\Delta t = 4$, $5$, and $6$~min.

In order to validate the calculation of the ice thickness using this model, we can compare the results obtained from 
the THz spectral data to the thickness calculation performed with the well established laser interference techniques described in Section~\ref{fringe}. 
The good agreement between the two methodologies confirm the validity of the present analysis.

\subsection{Laser interference technique}
\label{fringe}

In the adopted experimental configuration, a He-Ne laser beam (${\lambda}$ = 632.8 nm) is directed toward the sample and
reflected at near normal incidence both by the vacuum-sample and sample-substrate interfaces. The reflected beam is detected
by an external diode detector. It is possible to follow the accretion of the ice film by looking at the interference curve (intensity vs. time) 
of the reflected laser beam. 
Further details on the laser interference technique can be found in \cite{2016A&A...594A..80U}. \footnote{A publicly-available 
interface can be found at http://www.oact.inaf.it/spess/ to calculate the refractive index of the ice sample and derive the
theoretical interference curve from the amplitude of the experimental curve. The sample thickness is obtained by comparing
the two curves and using a procedure described in a document available at this web page.}

The results obtained with the two techniques have been compared. The data on the accretion of the ice vs. time obtained from the analysis of the interference curve, 
measured at the early stage of the deposition process, are in good agreement with the data obtained from the THz spectra.
In addition, by using the laser interference technique, we obtain for the CO ice a refractive index $n_{\rm CO} = 1.27$ that is close to the value obtained by using the THz technique.
The good agreement between the two methodologies confirm the validity of the present analysis.

\subsection{Reconstruction of the THz response}
\label{results}

Finally, from the first estimation of the ice thickness and complex refractive index, it is possible to reconstruct the THz dielectric response of ice. 
The reconstruction procedure is reported in Appendix~\ref{App-B}, while the results are summarized in Fig.~\ref{thick}. 
Panels (b) and (c) show the growth of the two CO ice layers versus the deposition time, considering different deposition steps of $\Delta t_{dep} = $ 4, 5, and 6 min.

\begin{figure}[h!]
    \centering
    \includegraphics[width=9.cm]{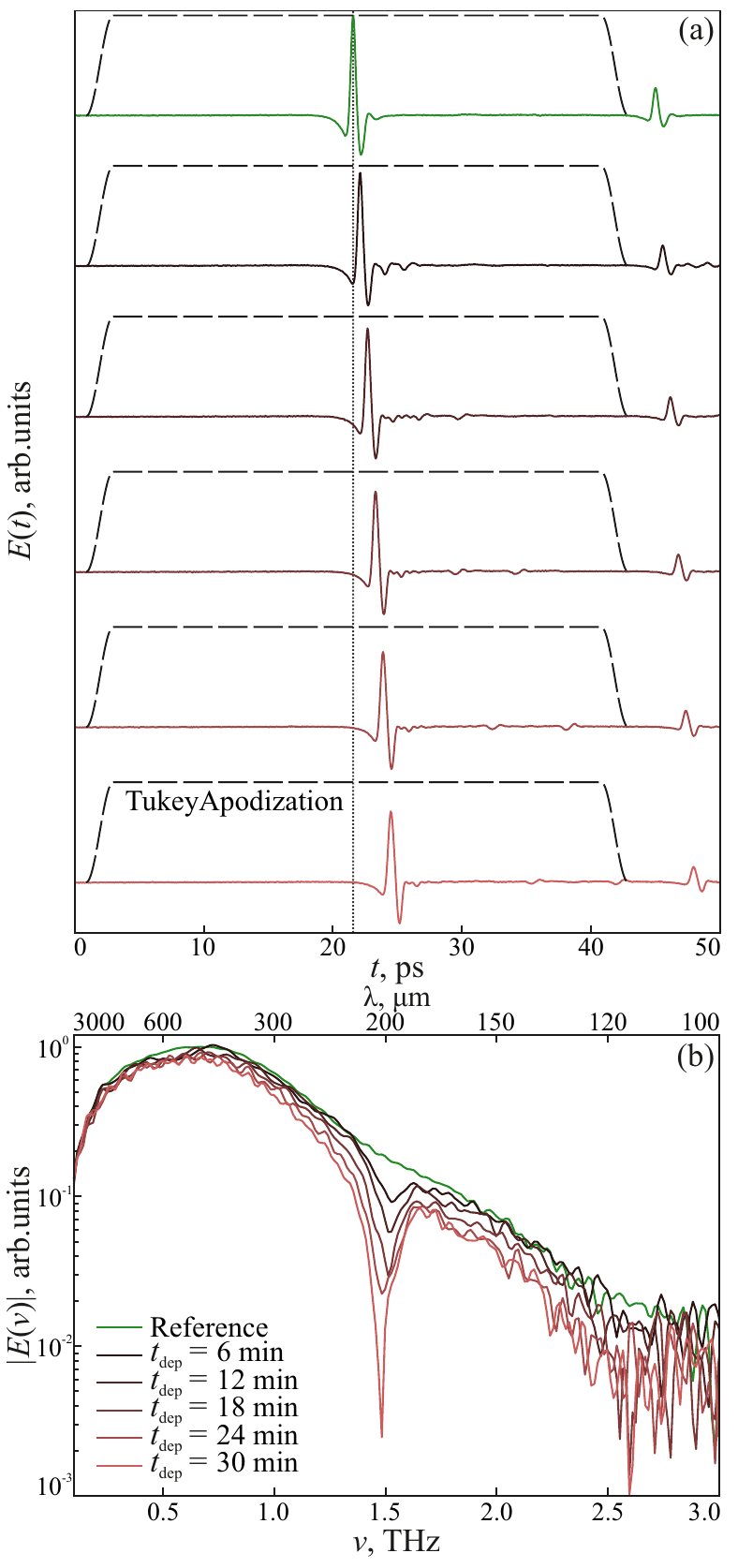}
    \caption{Evolution of the THz pulse and its spectra during the CO ice deposition. (a)~Reference waveform
    $E\left(t\right)$ transmitted through the cryostat with the empty substrate (green),
    and sample waveforms (black to light red) transmitted through the substrate with the CO ice deposited on its surfaces.
    (b)~Fourier spectra $| E \left( \nu \right) |$ of the reference and sample THz waveforms calculated with the use of
    Tukey apodization. The waveforms in (a) and spectra in (b) correspond to different values of the total deposition
    time $t_{\rm dep}$ (indicated), with the deposition intervals of $\Delta t_{\rm dep} = 6$~min.}
    \label{spectra}
\end{figure}

The recorded THz waveforms and their Fourier spectra are presented in Fig.~\ref{spectra} for the optical substrate (used as
a reference) and CO ice samples, at increasing thicknesses. In panel (a) the waveform $E \left( t \right)$ is shown for
the reference (green) and five subsequent deposition steps (black to light red) of approximately 0.45 mm total thickness for
each step. The thickness reached after the total deposition time is approximately 2.3~mm, split in two ice layers of
$\approx0.85$~mm and $\approx1.45$~mm on top of each side of the substrate.
A small source of inhomogeneity can be ascribed to the position of the pipe connected to the pumping system, which is located in a lateral position of the vacuum chamber with respect to the cold substrate.

In the THz spectrum of CO ice, we observe a Lorenz-like resonant peak centered near 1.5 THz (50~cm$^{-1}$) and
a second blurred feature close to 2.5 THz (83~cm$^{-1}$), masked by the sharp bands produced by the atmospheric water
contamination in the spectrometer sample compartment. 
The highest frequency accessible in our setup is presently limited by the strong absorption features  of residual water and carbon dioxide in the spectrometer case. 
We plan to change the current setup to an evacuated case in order to get rid of the contamination from the residual atmosphere, and we expect to extend the accessible frequency range up to 4 THz.

The estimated deposition rate for these deposition conditions is $\approx0.05$~mm/min for layer I and $\approx0.03$~mm/min
for layer II. These values are in a reasonable agreement with the results obtained employing the laser technique, where the
deposition rate is calculated to be 0.02~mm/min.
This agreement validates our hypothesis that the ice structure of thick ices does not differ significantly from the structure of thin ices, growing homogeneously over time during the deposition. 
The calculated optical properties are independent from the total thickness of the ice sample, allowing us to relate the laboratory data to the astrophysical ice conditions.

\begin{figure}[h!]
    \centering
    \includegraphics[width=9.cm]{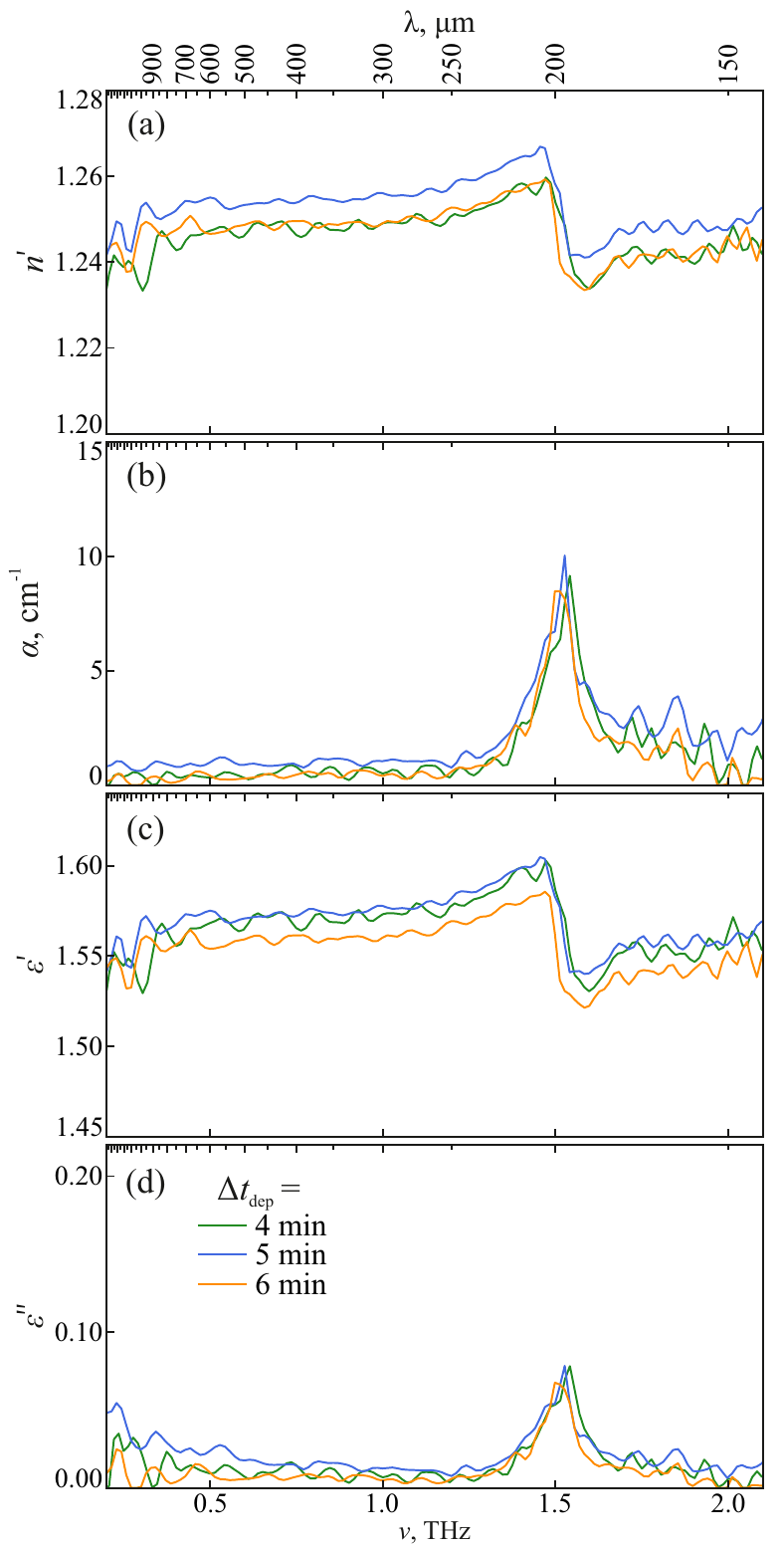}
    \caption{Optical properties of CO ice. (a)~Real part of the refractive index, (b)~amplitude absorption coefficient,
    (c)~real and (d)~imaginary parts of the dielectric permittivity (see Eqs.~1 and 2).
    For all deposition intervals, the dielectric curves demonstrate the existence of a Lorenz-like absorption
    peak, centered near $1.5$~THz and featuring similar bandwidth.
    Distortions of the results seen at frequencies below $0.3$~THz (such as an oscillatory
    character of $n$ and $\varepsilon'$ as well as an increase of $\varepsilon''$ with decreasing frequency)
    are due to diffraction effects (see Sec.~2.2).}
    \label{prop}
\end{figure}

Figure~\ref{prop} shows the determination of CO ice parameters, which are the refractive index (a), the amplitude absorption
coefficient (b), and both the real (c) and imaginary (d) parts of the complex dielectric permittivity -- see
Eqs.~\eqref{EQ:ComplexRefractiveIndex} and ~\eqref{EQ:ComplexDielectricPermittivity}.

\section{Discussion}
\label{disc}

A benefit of the direct reconstruction of the optical properties of the ices, provided by THz-TDS, is the detection of the
frequency dependent amplitude and phase of the waveform in a broad frequency range as a result of a single measurement.
These data eliminate the need of using the Kramers-Kronig relations for the reconstruction of the optical properties,
excluding additional distortion of the experimental data by edge effects, which frequently appears as a result of the Hilbert
integral transformation. This is of particular importance when dealing with broadband spectral kernels which are
usually present even when operating at low temperature.

We could also compare the refractive index of CO ice at THz frequencies with that previously calculated in the mid-IR range,
from \citet{AstroPhysJSupplSer.86.2.713.1993}, \citet{1997A&A...328..649E}, and \citet{CO_Baratta_1998} 
for a CO ice deposited at 10~K. However, a direct comparison between refrative index values found by different authors is not straightforward. 
As discussed by \citet{2005A&A...435..587L} and \citet{2017A&A...608A..81B}, the density and in turn the refractive index of an ice sample 
could strongly depend on the experimental conditions such as temperature, growth angle, and deposition rate.
We plan to test the effect of the change in the deposition conditions on the ice structure and check if this change will affect the optical constants. The results will be published in a forthcoming paper.

We compare the spectroscopic signature of the CO ice in our experiments with previous studies available in the literature. Data on far-IR spectra of solid CO were reported
by \citet{1966JChPh..45.4359A} and \citet{1967JChPh..46.3991R}. These studies investigated the absorption spectra of
amorphous CO, deposited at 10 $\mathrm{K}$ on a crystalline quartz substrate, between 250 and 30~cm$^{-1}$. Two bands are
visible at 50 and 83~cm$^{-1}$ (1.5 and 2.5 THz). The spectral features observed in our experiments are in excellent
agreement with these data, even though the 2.5 THz feature is masked by atmospheric water bands in our setup. The THz-TDS
technique has a low sensitivity, requiring very thick ice layers to be detectable.
Using our fast deposition rate the MIR vibrational bands of CO were strongly saturated within the first 30 seconds of deposition. 
We did some preliminary test on the ice growing using the vibrational bands in the NIR range and we were able to follow the ice growing up to ca. 160 $\mu$m thickness. 
Also in this case the ice growing is constant and linear in time during deposition, but the error associated with the thickness estimate using the NIR band is large (20-30\%).

It is not surprising, then, that in the data reported by \citet{C3FD00154G}, on THz and mid-IR spectroscopy of
interstellar ice analogs, the CO ice absorption band in the THz region has not been observed. The ice thickness in these
experiments, in fact, has been estimated to be in all cases less than 10~$\mu$m. In our case, the minimum thickness required
to be able to observe the 1.5~THz feature is estimated to be of the order of hundreds ${\mu}$m.

\begin{figure}[h!]
    \centering
    \includegraphics[width=9.cm]{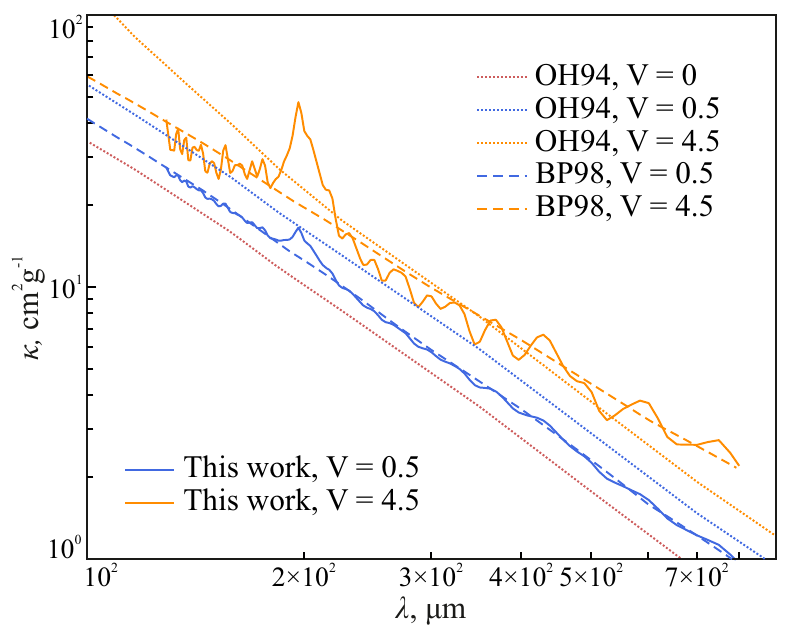}
    \caption{Calculated and reference opacities of astrophysical dust with CO ice and ice mixtures as a function of the wavelength. Dotted lines labelled with OH94 refer to bare grains and ice mixtures by \citet{1994A&A...291..943O}, 
    dashed lines with BP98 to CO ice by \citet{CO_Baratta_1998}, solid lines to the CO data by the present work. $V$ indicates the volume ratio between refractory core and ice mantle, 
    for which we follow \citet{1994A&A...291..943O}, where $V = 0$ (black) is the bare grain, $V = 0.5$ (blue) thin ice, and $V = 4.5$ (orange) thick ice. 
    See text for additional details.}
    \label{kappa}
\end{figure}

The data obtained are then employed to calculate the dust opacity for a given grain size distribution, as reported for example in 
\citet{1994A&A...291..943O} (see Appendix~\ref{App-C} for additional details on the method we developed to reproduce their results).
Following their approach, we report in Fig.~\ref{kappa} the calculated dust opacities assuming different ice coatings and different experimental data for the optical constants of ices.
Data from \citet{1994A&A...291..943O} are labelled OH94 and reported for bare grains (green dotted line), thin (blue dotted), and thick ice mantles (orange dotted).
The labels $V$ = 0, 0.5, and 4.5 indicate the volume ratio between the core of refractory material and the ice mantle (see Appendix~\ref{App-C}).
Conversely to the present work, \citet{1994A&A...291..943O} assume a H$_2$O:CH$_3$OH:CO:NH$_3 = $ 100:10:1:1 mixture ice mantle composition, 
i.e. water-based, with a minor amount of methanol, carbon monoxide, and ammonia. As described in detail in Appendix~\ref{App-C}, following the same procedure as in \citet{1994A&A...291..943O}, 
we extend the real and the imaginary parts of the refractive index from \citet{AstroPhysJSupplSer.86.2.713.1993} to longer wavelengths and we include spherical carbonaceous impurities.

The dielectric functions found by our experiments refer to pure CO ice, that explains the differences in the opacities shown in Fig.~\ref{kappa} 
calculated in the 100 -- 1000~$\mu$m range (solid lines). To have a more relevant comparison, we calculated the opacity using the same grain distribution and refractory materials as in 
\citet{1994A&A...291..943O} but with the optical values of CO ice coating extrapolated from \citet{CO_Baratta_1998} (labelled BP98), where the refractive index of CO ice 
deposited at 12~K is calculated from the spectrum recorded in the 4400 -- 400~cm$^{-1}$ (2.27 -- 25~$\mu$m) infrared spectral range. 
Their opacities (Fig.~\ref{kappa}, dashed lines) show an agreement with our results, except for the contribution due to the presence of a CO ice absorbing feature 
at approximately 200~$\mu$m, which is absent in the extrapolated data, as expected.

When we compare the refractive index, we note that the real part by BP98 ($n'$ = 1.28) is reasonable close to our data ($n'$ = 1.24), 
while for the imaginary part the discrepancy is very large, since $n''$ is completely determined by the absorption feature at 200 microns that is out of the range investigated 
by \citet{CO_Baratta_1998} (see Figure~\ref{prop}). 
It is worth to mention that \citet{1966JChPh..45.4359A} and \citet{1967JChPh..46.3991R} report a CO absorption feature at 2.5 THz, 
which is not visible in our spectra because it is masked by the atmospheric water contamination. This feature should further decrease the actual value of $n'$ 
at higher frequencies (above 2.5 THz). If we could measure the 2.5 THz feature and extend the calculation of the optical properties to this value, we would probably obtain a lower value of $n'$, 
slightly increasing the discrepancy with the data by \citet{CO_Baratta_1998}.
The reconstruction of the THz dielectric response of ices without the use of Kramers-Kronig relations, which is provided by THz pulsed spectroscopy, 
can provide an independent methodology to determine the optical properties of ice samples and validate the previous studies.

Since to compute the opacity both the real and the imaginary part of the dielectric constant are employed, we can infer from the calculated opacity curve that the imaginary part, 
which shows the biggest difference from the data presented by \citet{CO_Baratta_1998}, is not playing a major role in the determination of the opacity. 
This conclusion might be different for other absolute values or in a different spectral range.

\begin{table}[ht!]
  \caption[]{Opacities calculated at selected wavelengths and parameters of the fitting function $\kappa=\kappa_0 \left(\lambda / \lambda_0 \right)^{\beta}$, where $\lambda_0$ = 1 $\mu$m, for the two models with different volume ratios 
  (cf.~solid lines in Fig.~\ref{kappa}). $\kappa_0$ is in units of $10^5$~cm$^2$~g$^{-1}$.\label{kappatab}}
\begin{center}
    \begin{tabular}{lrr}
        \hline
        $\lambda/\mu$m & $\kappa_{V=0.5}$ & $\kappa_{V=4.5}$\\
        \hline
        200 & 14.94 & 38.12\\
        250 & 8.16 & 12.58\\
        350 & 4.28 & 6.28\\
        500 & 2.27 & 4.13\\
        800 & 0.97 & 2.22\\
        \hline
        \hline
        $\kappa_0/10^5$ & 2.28 & 1.28\\
        $\beta$ & -1.85 & -1.65\\
        \hline
    \end{tabular}
\end{center}
\end{table}

Values of opacities for some selected wavelengths are also given in Table~\ref{kappatab}.
The motivation for the choice of CO ice arises from the need of investigating the ice mantle properties for sources in which drastic CO depletion is expected, 
such as pre-stellar cores or protoplanetary disks midplanes \citep[e.g.][]{1999ApJ...523L.165C,2003A&A...408..981P}.
In these cases, an ice mantle rich in CO can be formed, and it will influence the optical properties of the dust grains. 
Thus, it is interesting to compare how the opacities change when ice mantles with diverse chemical composition are present.
The common moleculat species in astrophysical ices, such as H$_2$O, CO$_2$, NH$_3$ and possibly N$_2$, present absorption features in the FIR range. 
Therefore, the study of the influence of the spectroscopic features on the opacity of the ice mantles is important and we plan to extend this study to other pure ices and ice mixtures, 
which could be representative of the different molecular ice compositions in various astrophysical environments.

\section{Conclusions}
\label{conc}

In this work we have presented a study on the optical properties of solid CO at temperature and pressure conditions
significant for astrophysical applications. While previous data in  the mid-IR frequency range are available in the
literature, to the best of our knowledge, this study is the first to provide the complex refractive index and complex
dielectric permittivity of CO ice in the THz range.

We have shown that the ability of THz-TDS to measure both the amplitude and the phase information about the transmitted
pulse provides direct reconstruction of the complex dielectric function of ices without the use of the Kramers-Kronig
relations. The THz spectral features of ices can have a large bandwidth, such as the CO absorption line at $1.5$~THz. 
In this case, an implementation of the Kramers-Kronig relations (i.e., the Gilbert transform), relying only on the power
transmission/reflection spectrum, could lead to edge effects and resulting distortion of the dielectric response. Such
distortions are of particular importance when a spectral feature of the sample is located near the border of the spectral
operation range. Our results justify that THz-TDS setup is an appropriate instrument for accurate measurements of dielectric
properties of ices at THz frequencies.

These results have been used to calculate the opacities of the dust grains covered by a CO ice layer.
Discrepancies with currently available opacities suggest that measurements like those presented here are needed 
to provide a better interpretation of dust continuum emission, including dust and gas mass estimates.
In addition, they will provide further insight on the radiative transfer processes based on ice analogs optical and physical properties.

\begin{acknowledgements}
The authors acknowledge the anonymous referee for providing very useful comments which significantly improved the quality of the manuscript.

Also, the authors acknowledge Mr. Christian Deysenroth for the very valuable contribution in the designing and development of the
experimental setup.

We would like to thank Volker Ossenkopf for fruitful discussion to offering insight into the dust modelling methodology.

The work of Arsenii A. Gavdush on alignment of the THz-TDS setup and digital processing of the THz waveforms was supported
by the Russian Foundation for Basic Research (RFBR), Project \#$18$-$32$-$00816$.

The work of Gennady A. Komandin on solving the inverse problem of THz time-domain spectroscopy was supported by the Russian
Science Foundation (RSF), Project \#$18$-$12$-$00328$.

Tommaso Grassi acknowledges the support by the DFG clusterof excellence “Origin and Structure of the Universe” (http://www.universe-cluster.de/).  
This work was funded by the Deutsche Forschungsgemeinschaft (DFG, German Research Foundation) - Ref no. FOR 2634/1 ER685/11-1. 

This work has been partially supported by the project PRIN-INAF 2016, The Cradle of Life - GENESIS-SKA 
(General Conditions in Early Planetary Systems for the rise of life with SKA).

\end{acknowledgements}

\begin{appendix}
\section{THz-TDS optics}
\label{App-A}

In this section we provide a detailed description of the configuration of the laser beam inside the optical compartment of
the setup, already shown in Fig.~\ref{TDS}. The power of the laser beam is attenuated and successively divided into
equivalent channels. Thus, the antenna-emitter is pumped and the antenna-detector is probed with an equal average power of
about $20$~mW. The optical delay between the pump and probe beams is varied using a double-pass linear mechanical delay
stage from Zaber with the travel range of $101.6$~mm and the positioning accuracy of $<3$~$\mu$m. The THz radiation
undergoes $10$~kHz electrical modulation in order to synchronously detect the THz amplitude using a lock-in detection
principle.

The THz beam emitted by the photoconductive antenna is collimated by an integrated HRFZ-Si hemispherical lens and then
focused on a substrate window using a polymethylpentene (TPX) lens with the focal length of 67~mm and the diameter of 25~mm.
After passing through the cryostat vacuum chamber, it is collimated by an equal TPX lens in the direction of the
antenna-detector. Finally, the THz beam is focused onto the photoconductive gap of the antenna-detector by an equal
integrated HRFZ-Si hemispherical lens. In our measurements, during the waveform detection, we used a time-domain stride of
$50$~fs (it allows for satisfying the Whittaker-Nyquist-Kotelnikov-Shannon sampling theorem
\citep{TransAmerInstElectrEng.47.2.617.1928}), at time-domain window size of 100~ps and an averaging time of 0.1~sec at each
time-domain step with no waveform averaging.

The signal measured at the antenna-detector is recorded in the time domain $E \left( t \right)$ and converted into its
Fourier-spectrum $E \left( \nu \right)$ via 
\begin{equation}
    E \left( \nu \right) = \frac{1}{\sqrt{2 \pi}} \int_{-\infty}^{+\infty} E \left( t \right) e^{ - i 2 \pi \nu t } d t,
    \label{EQ:FourierTransform}
\end{equation}
where $t$ and $\nu$ stand for time and frequency.

\section{Reconstruction of the terahertz dielectric permittivity}
\label{App-B}

Eqs.~\eqref{Eq:FirsAssumtionForTheRefractiveIndex} and~\eqref{Eq:FirsAssumtionForTheThicknesses} are obtained as follows. 
We defined the complex amplitude of electromagnetic wave $E_\mathrm{0}=\left|E_\mathrm{0}\right|\exp\left(i\varphi_\mathrm{0}\right)$, 
which interacts either with a bare substrate (which we use as reference), or with two ice layers (when detecting the sample waveform). 
The pulses complex amplitudes depend on the initial complex amplitude of electromagnetic wave $E_\mathrm{0}$, as well as on  
ices and reference layers, which define the THz-wave reflection and transmission at the interfaces ($R$-operators and $T$-operators, 
respectively) and its absorption and phase delays in a bulk material ($P$-operators).

Then, we define the amplitudes of the ballistic reference pulse as follows:
\begin{equation}
    \begin{split}
        E_\mathrm{R,b}  =&
                       ~E_\mathrm{0}     P_\mathrm{0}    \left( l_\mathrm{0} - l_\mathrm{Si} \right) \,
                   T_\mathrm{0,Si} \,
                   P_\mathrm{Si}   \left( l_\mathrm{Si} \right) \,
                   T_\mathrm{Si,0} \\
                    = & \left|E_\mathrm{R,b}\right|\exp\left(i\varphi_\mathrm{R,b}\right),
    \end{split}
    \label{EQ:BallisticRefPulse}
\end{equation}
where $l$ is the thickness of the medium, and the symbols 0 and Si indicates respectively the vacuum and the HRFZ-Si,

the ballistic sample pulse as follows:
\begin{equation}
    \begin{split}
        E_\mathrm{S,b}=&
                       ~E_\mathrm{0} P_\mathrm{0} \left(l_{0} - l_{\rm Si} - l_{\rm CO,I} - l_{\rm CO,II} \right)
                       T_\mathrm{0,CO}
                       P_\mathrm{CO} \left( l_\mathrm{CO,I} \right) \\
                       &\times T_\mathrm{CO,Si}
                       P_\mathrm{Si} \left( l_\mathrm{Si} \right)
                       T_\mathrm{Si,CO}
                       P_\mathrm{CO}\left( l_\mathrm{CO,II} \right)
                       T_\mathrm{CO,0} \\
                       = & \left|E_\mathrm{S,b}\right|\exp\left(i\varphi_\mathrm{S,b}\right),
    \end{split}
    \label{EQ:BallisticSamPulse}
\end{equation}
where $l_{\rm CO,I}$ and $l_{\rm CO,II}$ are defined as in Section~\ref{model},

and the two satellite sample pulses as follows:
\begin{equation}
    \begin{split}
        E_\mathrm{S,1s}=&
                       ~E_\mathrm{0} P_\mathrm{0} \left(l_{0} - l_{\rm Si} - l_{\rm CO,I} - l_{\rm CO,II} \right)
                       T_\mathrm{0,CO}
                       P_\mathrm{CO} \left( l_\mathrm{CO,I} \right) \\
                       &\times T_\mathrm{CO,Si}
                       P_\mathrm{Si} \left( l_\mathrm{Si} \right)
                       T_\mathrm{Si,CO}
                       P_\mathrm{CO}\left( l_\mathrm{CO,II} \right)
                       T_\mathrm{CO,0}\\
                       &\times R_\mathrm{CO,Si}
                                P_\mathrm{CO}^{2} \left( l_\mathrm{CO,I} \right)
                                R_\mathrm{CO,0} \\
                       = & \left|E_\mathrm{S,1s}\right|\exp\left(i\varphi_\mathrm{S,1s}\right), \\
        E_\mathrm{S,2s}=&
                       ~E_\mathrm{0} P_\mathrm{0} \left(l_{0} - l_{\rm Si} - l_{\rm CO,I} - l_{\rm CO,II} \right)
                       T_\mathrm{0,CO}
                       P_\mathrm{CO} \left( l_\mathrm{CO,I} \right) \\
                       &\times T_\mathrm{CO,Si}
                       P_\mathrm{Si} \left( l_\mathrm{Si} \right)
                       T_\mathrm{Si,CO}
                       P_\mathrm{CO}\left( l_\mathrm{CO,II} \right)
                       T_\mathrm{CO,0}\\
                       &\times R_\mathrm{CO,0}
                                P_\mathrm{CO}^{2} \left( l_\mathrm{CO,II} \right)
                                R_\mathrm{CO,Si}\\
                       = & \left|E_\mathrm{S,2s}\right|\exp\left(i\varphi_\mathrm{S,2s}\right),
    \end{split}
    \label{EQ:SatSam}
\end{equation}
which are clearly observed in reference and sample TDS waveforms in Fig.~\ref{thick}~(a).

Then, by neglecting the phase changes during the reflection at the interfaces of absorbing media 
(these changes are weak since we study rather low-absorbing dielectric materials) as well as the distortion of optical pulses due to dispersion of material parameters 
(the dispersion of HRFZ-Si is very low, while the dispersion in ice is negligible due to its small thickness), we calculated the phases of these pulses:
\begin{equation}
    \begin{split}
        \varphi_\mathrm{R,b}=&
                       ~\frac{ 2 \pi \nu } { c } \left(n_\mathrm{0}\left(l_\mathrm{0}-l_\mathrm{Si}\right) + n_\mathrm{Si}l_\mathrm{Si} \right) \\
                       = & -\frac{ 2 \pi \nu } { c } \left(l_\mathrm{0}-l_\mathrm{Si}\left(n_\mathrm{Si}-1\right)\right), \\
       \varphi_\mathrm{S,b}=&
                       ~\frac{ 2 \pi \nu } { c } \left(n_\mathrm{0}\left(l_\mathrm{0}-l_\mathrm{Si}-l_\mathrm{CO,I}-l_\mathrm{CO,II}\right) + n_\mathrm{Si}l_\mathrm{Si} \right.\\
                       &\left.+n_\mathrm{CO}\left(l_\mathrm{CO,I}+l_\mathrm{CO,II}\right) \right) \\
                       = & -\frac{ 2 \pi \nu } { c } \left(l_\mathrm{0}-l_\mathrm{Si}\left(n_\mathrm{Si}-1\right)-\left(l_\mathrm{CO,I}+l_\mathrm{CO,II}\right)\left(n_\mathrm{CO}-1\right)\right),\\
       \varphi_\mathrm{S,1s}=&
                       ~\frac{ 2 \pi \nu } { c } \left(n_\mathrm{0}\left(l_\mathrm{0}-l_\mathrm{Si}-l_\mathrm{CO,I}-l_\mathrm{CO,II}\right) + n_\mathrm{Si}l_\mathrm{Si} \right.\\
                       &\left.+ n_\mathrm{CO}\left(3l_\mathrm{CO,I}+l_\mathrm{CO,II}\right) \right) \\
                       = & -\frac{ 2 \pi \nu } { c } \left(l_\mathrm{0}-l_\mathrm{Si}\left(n_\mathrm{Si}-1\right)-\left(l_\mathrm{CO,I}+l_\mathrm{CO,II}\right)\left(n_\mathrm{CO}-1\right)\right.\\
                       &\left.- 2l_\mathrm{CO,I}n_\mathrm{CO}\right), \\
       \varphi_\mathrm{S,2s}=&
                       ~\frac{ 2 \pi \nu } { c } \left(n_\mathrm{0}\left(l_\mathrm{0}-l_\mathrm{Si}-l_\mathrm{CO,I}-l_\mathrm{CO,II}\right) + n_\mathrm{Si}l_\mathrm{Si} \right.\\
                       &\left.+n_\mathrm{CO}\left(l_\mathrm{CO,I}+3l_\mathrm{CO,II}\right) \right) \\
                       = & -\frac{ 2 \pi \nu } { c } \left(l_\mathrm{0}-l_\mathrm{Si}\left(n_\mathrm{Si}-1\right)-\left(l_\mathrm{CO,I}+l_\mathrm{CO,II}\right)\left(n_\mathrm{CO}-1\right)\right.\\
                       &\left.- 2l_\mathrm{CO,II}n_\mathrm{CO}\right). \\
    \end{split}
    \label{EQ:SignalPhases}
\end{equation}
where $n_\mathrm{0}$ and $n_\mathrm{Si}$ are defined as in Section~\ref{model} and $n_\mathrm{CO}$ is the  refractive index of the CO ice.

These phases are used to calculate the time delays between the pulses, which are indicated in Fig.~\ref{thick}~(a):
\begin{equation}
    \begin{split}
    \delta t_\mathrm{01}=&
                       ~\frac{\varphi_\mathrm{S,b}-\varphi_\mathrm{R,b}}{2 \pi \nu} = \frac{ n_\mathrm{CO}-1 } { c } \left(l_\mathrm{CO,I}+l_\mathrm{CO,II}\right), \\
    \delta t_\mathrm{12}=&
                       ~\frac{\varphi_\mathrm{S,1s}-\varphi_\mathrm{S,b}}{2 \pi \nu} = \frac{ 2n_\mathrm{CO} } { c } l_\mathrm{CO,I}, \\
    \delta t_\mathrm{13}=&
                       ~\frac{\varphi_\mathrm{S,2s}-\varphi_\mathrm{S,b}}{2 \pi \nu} = \frac{ 2n_\mathrm{CO} } { c } l_\mathrm{CO,II}. \\
    \end{split}
    \label{EQ:SignalTimeDelay}
\end{equation}
Solving this system of equations yields Eqs.~\eqref{Eq:FirsAssumtionForTheRefractiveIndex} and~\eqref{Eq:FirsAssumtionForTheThicknesses}.

Figure~\ref{spectra}~(b) shows the Fourier spectra $E \left( \nu \right)$ of the reference and sample TDS waveforms. In
order to filter out the contribution of the satellite THz pulses (caused by the interference in the input and output windows
and the substrate), and to improve the analysis of the frequency-domain data, we apply equal apodization procedure (window filtering) to all waveforms,
\begin{equation}
    E_\mathrm{filt} \left( t \right) = E \left( t \right) H \left( t - t_\mathrm{0} \right),
    \label{EQ:Apodization}
\end{equation}
where $E \left( t \right)$ and $E_\mathrm{filt} \left( t \right)$ stand for the initial and filtered waveforms, $H \left( t \right)$ defines the apodization function,
$t_\mathrm{0}$ defines a position of the apodization filter towards the THz waveform,

\begin{equation}
    H \left( t \right) =
    \begin{cases}
        \frac12 + \frac12\cos \left[ \frac{2 \pi}{\omega} \left( \frac{t}{\tau} - \frac{ \omega} {2} \right)  \right],
        & 0 < \frac{t}{\tau} < \frac{\omega}{2}\,,\vspace{.2cm}\\
        1,
        & \frac{\omega}{2} < \frac{t}{\tau} < 1 - \frac{\omega}{2}\vspace{.2cm}\\
        \frac{1}{2} + \frac12\cos \left[ \frac{2 \pi}{\omega} \left(\frac{t}{\tau} - 1 + \frac{ \omega} {2} \right)  \right],
        & 1 - \frac{\omega}{2} < \frac{t}{\tau} < 1\,,
    \end{cases},
    \label{EQ:Apodization}
\end{equation}

is a Tukey apodization filter \citep{TukeyBook.1986} with the width $\tau$, and $\omega$ stands for a parameter of the
filter smoothness (for $\omega = 0$ the window has a rectangular form, for $\omega = 1$ it is the Hann window
\cite{ProcIEEE.66.1.51.1978}). As shown in Fig.~\ref{spectra}(a), we use the Tukey window centered at the maximum of the
reference THz waveform, with the smoothness parameter of $\omega = 0.1$ and the width of $40$~ps (which yields the
frequency-domain resolution of $0.025$~THz).

Let us consider the Fresnel formulas \citep{WolfBorn.Book.1980}, defining the THz wave amplitude reflection at (and
transmission through) the interface between media $m$ and $k$:

\begin{equation}
    \begin{split}
        R_{m,k} \left( \nu \right) &= \frac{n_{m}\left(\nu\right)-n_{k}\left( \nu \right)}{n_{m}\left(\nu\right)+n_{k}\left(\nu\right)},\\
        T_{m,k} \left( \nu \right) &= \frac{2 n _{m} \left( \nu \right) }{ n_{m} \left( \nu \right) + n_{k} \left( \nu \right) },
    \end{split}
    \label{EQ:FresnelFormulas}
\end{equation}

where $R_{m,k} \left( \nu \right)$ and $T_{m,k} \left( \nu \right)$ stand for coefficients of the complex amplitude
reflection and transmission, respectively, while $n_m(\nu) + i n_k(\nu)$ is the complex refractive index of the media. 
The relation between complex amplitudes of the THz wave right after the emitter ($z = 0$), $E_0 \left(
\nu\right)$, and at the position $z$ along the beam axis, $E \left( \nu, z \right)$, is given by

\begin{equation}
    E \left( \nu, z \right) = E_0 \left( \nu\right) \exp \left( - i \frac{ 2 \pi \nu } { c } n \left( \nu \right) z \right).
    \label{EQ:BouguerLambertBeerLaw}
\end{equation}

If the thicknesses and the refractive indexes of all layers are known, Eqs.~\eqref{EQ:FresnelFormulas}
and~\eqref{EQ:BouguerLambertBeerLaw} yield description of all peculiarities of the THz pulse interacting with
multilayer structures \citep{JApplPhys.115.19.193105.2014,IEEETransTerSciTech.5.5.817.2015,Gavdush2019}.

We derive the equations describing the complex amplitudes of the reference $E_\mathrm{R} \left( \nu \right)$ and sample
$E_\mathrm{S} \left( \nu \right)$ spectra, assuming only wavelets inside the Tukey apodization.

For the reference spectrum, we obtain
\begin{equation}
    E_\mathrm{R}/E_\mathrm{0}  =
                   P_\mathrm{0}    \left( l_\mathrm{0} - l_\mathrm{Si} \right) \,
                   T_\mathrm{0,Si} \,
                   P_\mathrm{Si}   \left( l_\mathrm{Si} \right) \,
                   T_\mathrm{Si,0},
    \label{EQ:ReferenceWaveform}
\end{equation}
where the indexes $\mathrm{0}$ and $\mathrm{Si}$ correspond to the free space and the HRFZ-Si medium; $l_\mathrm{0}$ and
$l_\mathrm{Si}$ are the total length of the THz beam path and the thickness of the HRFZ-Si substrate, respectively; 
$T_\mathrm{0,Si} \left( \nu \right)$ and $T_\mathrm{Si,0} \left( \nu
\right)$ are the transmission coefficients for the respective interfaces (between the free space and the HRFZ-Si),
defined by Eq.~\eqref{EQ:FresnelFormulas}; $P_\mathrm{0} \left( \nu, z \right)$ and $P_\mathrm{Si} \left( \nu, z \right)$
are operators describing the THz wave propagation in the free space and the HRFZ-Si, respectively, as given by the exponential factor in
Eq.~\eqref{EQ:BouguerLambertBeerLaw}.

As shown in Fig.~\ref{layers}, for the sample spectrum we take into account the contribution of the ballistic THz pulse (1)
and the satellite pulses (2 and 3), caused by the multiple THz wave reflection in the ice films. This yields the following
equation
\begin{equation}
    \begin{split}
        E_\mathrm{S}/E_\mathrm{0}=&
                       ~P_\mathrm{0} \left(l_{0} - l_{\rm Si} - l_{\rm CO,I} - l_{\rm CO,II} \right)
                       T_\mathrm{0,CO}
                       P_\mathrm{CO} \left( l_\mathrm{CO,I} \right) \\
                       &\times T_\mathrm{CO,Si}
                       P_\mathrm{Si} \left( l_\mathrm{Si} \right)
                       T_\mathrm{Si,CO}
                       P_\mathrm{CO}\left( l_\mathrm{CO,II} \right)
                       T_\mathrm{CO,0}\\
                       &\times \left[  1
                              + R_\mathrm{CO,Si}
                                P_\mathrm{CO}^{2} \left( l_\mathrm{CO,I} \right)
                                R_\mathrm{CO,0}\right.\\
                           &\left.+ R_\mathrm{CO,0}
                            P_\mathrm{CO}^{2} \left( l_\mathrm{CO,II} \right)
                            R_\mathrm{CO,Si}
                       \right],
    \end{split}
    \label{EQ:SampleWaveform}
\end{equation}
where the summation terms in the brackets correspond to the wavelets 1, 2 and 3; $l_\mathrm{CO,I}$ and $l_\mathrm{CO,II}$
stand for thicknesses of the first and second ice films;
$R_\mathrm{CO,Si} \left( \nu \right)$ and $R_\mathrm{CO,0} \left( \nu \right)$ are the transmission coefficients for
the respective interfaces, see Eq.~\eqref{EQ:FresnelFormulas}; the definition of the remaining factors (P and T) is similar to that in Eqs.~\eqref{EQ:ReferenceWaveform}.
We point out that the complex amplitudes ($E$) and all factors ($T$, $R$, and $P$)
in Eq.~\eqref{EQ:ReferenceWaveform} and~\eqref{EQ:SampleWaveform} are frequency-dependent.

Equations~\eqref{EQ:ReferenceWaveform} and~\eqref{EQ:SampleWaveform} form a basis for the reconstruction of the THz
dielectric response of ices.

The reconstruction is performed via the following minimization procedure:
\begin{equation}
    n \left( \nu \right) = \mathrm{argmin}_{n \left( \nu \right)} \left[ \mathbf{\Phi} \left( n \left(\nu\right),\nu\right)\right],
    \label{EQ:ErrorFunctionalMinimization}
\end{equation}
where $\mathrm{argmin}$ is an operator which determines the minimum argument of the vector error functional
$\mathbf{\Phi}$. The latter is formed from the complex theoretical $T_\mathrm{Th}$ and experimental $T_\mathrm{Exp}$
transfer functions,
\begin{equation}
    \mathbf{\Phi} \left( n \left( \nu \right), \nu \right) =
    \left( \begin{array}{c}
\left| T_\mathrm{Th} \left( n, \nu \right) \right| - \left| T_\mathrm{Exp} \left( \nu \right) \right| \\
\phi \left[ T_\mathrm{Th} \left( n, \nu \right) \right] - \phi \left[ T_\mathrm{Exp} \left( \nu \right) \right]
\end{array} \right),
    \label{EQ:ErrorFunctionalFormulation}
\end{equation}
where $|...|$ and $\phi \left[ ... \right]$ are, respectively, the absolute values and the phases of the complex
functions.

We define the theoretical transfer function $T_\mathrm{Th}$ as the sample spectrum (Eq.~\ref{EQ:SampleWaveform}) normalized
by the reference spectrum (Eq.~\ref{EQ:ReferenceWaveform}),
\begin{equation}
    \begin{split}
        T_\mathrm{Th} = &~\frac{ T_\mathrm{0,CO}
                               T_\mathrm{CO,Si}
                               T_\mathrm{Si,CO}
                               T_\mathrm{CO,0}}
                             { T_\mathrm{0,Si}
                               T_\mathrm{Si,0}} \\
                        &\times P_\mathrm{CO} \left(   l_\mathrm{CO,I} + l_\mathrm{CO,II} \right)
                         P_\mathrm{0}  \left( - l_\mathrm{CO,I}  - l_\mathrm{CO,II}      \right) \\
                         &\times\left[  1
                              + R_\mathrm{CO,Si}
                                P_\mathrm{CO}^{2} \left( l_\mathrm{CO,I} \right)
                                R_\mathrm{CO,0} \right.\\
                             &\left.+ R_\mathrm{CO,0}
                                P_\mathrm{CO}^{2} \left( l_\mathrm{CO,II} \right)
                                R_\mathrm{CO,Si}
                           \right].
    \end{split}
    \label{EQ:TheoreticalTransferFunction}
\end{equation}
Considering all reflection and transmission operators in Eq. (B.9), we note that $T_\mathrm{Th}$ depends only on the
refractive index of the HRFZ-Si substrate $n_\mathrm{Si}$, which is known \textit{a priori}, as well as on the parameters of
the CO ice to be determined; it excludes contribution of several factors, such as the unknown complex amplitude of the TDS
source $E_\mathrm{0} \left( \nu \right)$, the unknown total length of the THz beam path $l_\mathrm{0}$, and, finally, the
thickness of the HRFZ-Si substrate $l_\mathrm{Si}$, which is known, too, but can slightly vary owing to angular deviations
of the substrate during the vacuum chamber assembling.

The experimental transfer function $T_\mathrm{Exp}$ is calculated in a similar manner, relying on the Fourier spectra of the
experimental sample $E_\mathrm{S}$ and reference $E_\mathrm{R}$ waveforms (after applying the Tukey apodization),
\begin{equation}
    T_\mathrm{Exp} = \frac{E_\mathrm{S}}{E_\mathrm{R}}.
    \label{EQ:ExperimentalTransferFunction}
\end{equation}
Notice that all the functions and operators in the theoretical and experimental transfer functions are frequency-dependent,
and both transfer functions take into account only the ballistic pulses of the reference and sample waveforms, as well as
the first and second ice-related satellite pulses of the sample waveform.

By introducing equal confidence intervals for the refractive index $n'$ and the amplitude absorption coefficient
$\alpha$, as $[n'_\mathrm{init} - \Delta n', n'_\mathrm{init} + \Delta n']$ and $[0, \alpha_\mathrm{max}]$ with $\Delta n'
=0.25$ and $\alpha_\mathrm{max} = 15$~cm$^{-1}$, we use the Non-Linear Trust Region Approach \citep{SIAMJOptim.6.2.418.1996}
to reconstruct the THz dielectric response of the CO ice in the spectral operation range of $0.3$ to $2.0$~THz.

\section{Opacity model benchmark}
\label{App-C}

In order to verify the correctness of the machinery employed to calculate the opacity from the dielectric constants, we reproduce the results found by \citet{1994A&A...291..943O} in their Fig.~5, panels a-c, \emph{compact grains} case.
We describe here the methodology employed. The results described here can be reproduced by running \texttt{test\_04.py} from the publicly available code\footnote{\url{https://bitbucket.org/tgrassi/compute_qabs}, commit: 8c0812f}, 
while Fig.~\ref{kappa} from our paper can be reproduced with \texttt{test\_05.py}.

\subsection{Dielectric constants}
The complex dielectric functions $\eps'-i\eps''$ of the refractory components are taken from \citet{1992A&A...261..567O} (cool oxygen-rich silicates, their Fig.~10) 
and from \citet{1993A&A...279..577P} (amorphous carbon, their Table~1), while ice is assumed to be a H$_2$O:CH$_3$OH:CO:NH$_3 = $ 100:10:1:1 mixture at $10$~K from \citet{AstroPhysJSupplSer.86.2.713.1993} (their Table~2A).

For the aims of this work we need to extrapolate the ice data relative to the H$_2$O:CH$_3$OH:CO:NH$_3 = $ 100:10:1:1 mixture to longer wavelengths, as done by \citet{1994A&A...291..943O}. First we fit the last 45~data points\footnote{This number is chosen manually to select the 
$\lambda^{-1}$ decaying part of the data after the last available resonance.} of the imaginary part (i.e.~approximately $71-194$~$\mu$m) with a $f(\lambda)\propto\lambda^{-1}$ function (Ossenkopf, priv. comm.), 
and we use this to extrapolate $\eps''$ with 200~linearly-spaced wavelength points, in the range $71-800$~$\mu$m. To retrieve the real part of $\eps$ at each $\freq=2\pi c\lambda^{-1}$ point, we apply the 
Kramers-Kronig \citep[e.g, ][]{1983asls.book.....B} relation in the discrete form
\beq
 \eps'(\freq) = 1 + \frac{2}{\pi} \Omega(\freq)\,,
\eeq
with $\Omega(\freq)$ the discrete integral over the positive frequency ranges using the composite trapezoidal rule with the integrand $\frac{\eps''(\freq_i)\,\freq_i}{\freq_i^2 - \freq^2}$, 
and excluding $\freq$, where the denominator of the argument vanishes (i.e.~the Cauchy principal value of the corresponding finite integral).

After the extrapolation, the ice dielectric functions are modified by mixing spherical inclusions of amorphous carbon using the Bruggeman effective medium approximation \citep{1935AnP...416..636B}, with a volume filling fraction 
of $0.11$ and $0.013$ for the \emph{thin} and the \emph{thick} ice cases, respectively.

\subsection{Absorption coefficients}
The absorption efficiency $Q_{abs}(\lambda, a)$ is function of the wavelength and of the grain size $a$, and computed with the routine \textsc{bhmie.py} for the bare grains\footnote{\url{http://scatterlib.wikidot.com/mie}} 
and with \textsc{bhcoat.py} for the coated grains\footnote{Adapted here from the \textsc{Fortran} version at \url{http://scatterlib.wikidot.com/coated-spheres}.}, both from \citep{1983asls.book.....B}. In \citet{1994A&A...291..943O} there are three cases, 
\emph{bare} grains without ice, \emph{thin} ice, that has a volume ratio of $V=0.5$, and \emph{thick} ice, where $V=4.5$. The radius of the refractory core $a$ and the fraction $V$ determine the radius of the mantle $a_{coat} = a(V+1)^{1/3}$.

\subsection{Opacities}
With $Q_{abs}$ it is possible to retrieve the opacities averaged on the grain size distribution $\varphi(a)$ as
\beq
 \kappa(\lambda) = \frac{\pi}{C}{\int_{a_{min}^{coat}}^{a_{max}^{coat}} \varphi(a)\, a^2 Q(\lambda, a)\, \dd a}\,,
\eeq
where $a_{min}^{coat}$ to $a_{max}^{coat}$ is the range where the size distribution is valid \emph{including coating}, and
\beq
 C = \frac{4}{3}\pi\rho_0 \int_{a_{min}}^{a_{max}} \varphi(a)\,a^3 \dd a\,,
\eeq
where $a_{min}$ to $a_{max}$ is the range where the size distribution is valid, but considering \emph{only the refractory material}, i.e.~silicates or carbonaceous dust.

\citet{1994A&A...291..943O} assume  $a_{min}=5\times10^{-7}$~cm, $a_{max}=2.5\times10^{-5}$~cm, $\rho_0=2.9$~g~cm$^{-3}$ (silicates) and $\rho_0=2$~g~cm$^{-3}$ (amorphous carbon), and $\varphi(a) = a^{-3.5}$.

The total opacity is $\kappa_{tot}(\lambda) = 0.678\kappa_{Si}(\lambda)+0.322\kappa_{AC}(\lambda)$, where the two terms are respectively the silicates and carbonaceous opacities including ice coating, 
and the two coefficients are calculated from the volume ratio discussed in Sect.~3.1 of \citet{1994A&A...291..943O}. In particular, \citeauthor{1994A&A...291..943O} assume a volume ratio of the refractory components $V_{\rm AC} / V_{\rm Si} = 0.69$ (their Sect.~3.1), 
that can be converted into the corresponding mass ratio $M_{\rm AC} / M_{\rm Si} = 0.4758$ by using the relation $M_i = \rho_i V_i$, where $\rho_i$ are the bulk densities of the two refractory components 
(being in \citeauthor{1994A&A...291..943O} the opacity defined per unit mass of the refractory material), so that $\kappa_{tot} = (M_{AC} \kappa_{AC} + M_{Si} \kappa_{Si}) / (M_{AC} + M_{Si})$.

\end{appendix}

\bibliographystyle{aa} 

\begin{thebibliography}{43}
\expandafter\ifx\csname natexlab\endcsname\relax\def\natexlab#1{#1}\fi

\bibitem[{{Allodi} {et~al.}(2014){Allodi}, {Ioppolo}, {Kelley}, {McGuire}, \&
  {Blake}}]{2014PCCP...16.3442A}
{Allodi}, M.~A., {Ioppolo}, S., {Kelley}, M.~J., {McGuire}, B.~A., \& {Blake},
  G.~A. 2014, Physical Chemistry Chemical Physics (Incorporating Faraday
  Transactions), 16, 3442

\bibitem[{{Anderson} \& {Leroi}(1966)}]{1966JChPh..45.4359A}
{Anderson}, A. \& {Leroi}, G.~E. 1966, \jcp, 45, 4359

\bibitem[{Auston(1975)}]{Auston_TDS_1975}
Auston, D.~H. 1975, Applied Physics Letters, 26, 101

\bibitem[{Baratta \& Palumbo(1998)}]{CO_Baratta_1998}
Baratta, G.~A. \& Palumbo, M.~E. 1998, Journal of the Optical Society of
  America A: Optics and Image Science, and Vision, 15, 3076

\bibitem[{{Baratta} \& {Palumbo}(2017)}]{2017A&A...608A..81B}
{Baratta}, G.~A. \& {Palumbo}, M.~E. 2017, \aap, 608, A81

\bibitem[{{Bohren} \& {Huffman}(1983)}]{1983asls.book.....B}
{Bohren}, C.~F. \& {Huffman}, D.~R. 1983, {Absorption and scattering of light
  by small particles} (Wiley)

\bibitem[{Born \& Wolf(1980)}]{WolfBorn.Book.1980}
Born, M. \& Wolf, E. 1980, {Principles of Optics. 6th Edition} (UK: Pergamon
  Press)

\bibitem[{{Bruggeman}(1935)}]{1935AnP...416..636B}
{Bruggeman}, D.~A.~G. 1935, Annalen der Physik, 416, 636

\bibitem[{{Caselli} {et~al.}(1999){Caselli}, {Walmsley}, {Tafalla}, {Dore}, \&
  {Myers}}]{1999ApJ...523L.165C}
{Caselli}, P., {Walmsley}, C.~M., {Tafalla}, M., {Dore}, L., \& {Myers}, P.~C.
  1999, \apjl, 523, L165

\bibitem[{Coleman \& Li(1996)}]{SIAMJOptim.6.2.418.1996}
Coleman, T. \& Li, Y. 1996, SIAM Journal on Optimization, 6, 418

\bibitem[{{Dartois}(2006)}]{2006A&A...445..959D}
{Dartois}, E. 2006, \aap, 445, 959

\bibitem[{{Dutrey} {et~al.}(1998){Dutrey}, {Guilloteau}, {Prato}, {Simon},
  {Duvert}, {Schuster}, \& {Menard}}]{1998A&A...338L..63D}
{Dutrey}, A., {Guilloteau}, S., {Prato}, L., {et~al.} 1998, \aap, 338, L63

\bibitem[{{Ehrenfreund} {et~al.}(1997){Ehrenfreund}, {Boogert}, {Gerakines},
  {Tielens}, \& {van Dishoeck}}]{1997A&A...328..649E}
{Ehrenfreund}, P., {Boogert}, A.~C.~A., {Gerakines}, P.~A., {Tielens},
  A.~G.~G.~M., \& {van Dishoeck}, E.~F. 1997, \aap, 328, 649

\bibitem[{Gavdush {et~al.}(2019)Gavdush, Chernomyrdin, Malakhov, Beshplav,
  Dolganova, Kosyrkova, Nikitin, Musina, Katyba, Reshetov, Cherkasova,
  Komandin, Karasik, Potapov, Tuchin, \& Zaytsev}]{Gavdush2019}
Gavdush, A.~A., Chernomyrdin, N., Malakhov, K., {et~al.} 2019, Journal of
  Biomedical Optics, 24, 027001

\bibitem[{{Giuliano} {et~al.}(2014){Giuliano}, {Escribano},
  {Mart{\'i}n-Dom{\'e}nech}, {Dartois}, \& {Mu{\~n}oz
  Caro}}]{2014A&A...565A.108G}
{Giuliano}, B.~M., {Escribano}, R.~M., {Mart{\'i}n-Dom{\'e}nech}, R.,
  {Dartois}, E., \& {Mu{\~n}oz Caro}, G.~M. 2014, \aap, 565, A108

\bibitem[{{Giuliano} {et~al.}(2016){Giuliano}, {Mart{\'i}n-Dom{\'e}nech},
  {Escribano}, {Manzano-Santamar{\'i}a}, \& {Mu{\~n}oz
  Caro}}]{2016A&A...592A..81G}
{Giuliano}, B.~M., {Mart{\'i}n-Dom{\'e}nech}, R., {Escribano}, R.~M.,
  {Manzano-Santamar{\'i}a}, J., \& {Mu{\~n}oz Caro}, G.~M. 2016, \aap, 592, A81

\bibitem[{Griffiths \& de~Haseth(1986)}]{GriffithsFTIR_Book1986}
Griffiths, P. \& de~Haseth, J. 1986, {Fourier Transform Infrared Spectroscopy}
  (New York, NY, USA: John Wiley + Sons)

\bibitem[{Harris(1978)}]{ProcIEEE.66.1.51.1978}
Harris, F.~J. 1978, Proceedings of the IEEE, 66, 51

\bibitem[{Hudgins {et~al.}(1993)Hudgins, Sandford, Allamandola, \&
  Tielens}]{AstroPhysJSupplSer.86.2.713.1993}
Hudgins, D.~M., Sandford, S.~A., Allamandola, L.~J., \& Tielens, A. G. G.~M.
  1993, Astrophysical Journal Supplement Series, 86, 713

\bibitem[{Ioppolo {et~al.}(2014)Ioppolo, McGuire, Allodi, \&
  Blake}]{C3FD00154G}
Ioppolo, S., McGuire, B.~A., Allodi, M.~A., \& Blake, G.~A. 2014, Faraday
  Discuss., 168, 461

\bibitem[{Kawase {et~al.}(1996)Kawase, Sato, Taniuchi, \& Ito}]{PC_Kawase_1996}
Kawase, K., Sato, M., Taniuchi, T., \& Ito, H. 1996, Applied Physics Letters,
  68, 2483

\bibitem[{Kiessling {et~al.}(2013)Kiessling, Breunig, Schunemann, Buse, \&
  Vodopyanov}]{CW_THz_OPO_2013}
Kiessling, J., Breunig, I., Schunemann, P.~G., Buse, K., \& Vodopyanov, K.~L.
  2013, New Journal of Physics, 15, 105014

\bibitem[{Komandin {et~al.}(2013)Komandin, Chuchupal, Lebedev, Goncharov,
  Korolev, Porodinkov, Spektor, \& Volkov}]{IEEE_BWO_2013}
Komandin, G.~A., Chuchupal, S.~V., Lebedev, S.~P., {et~al.} 2013, IEEE
  Transactions on Terahertz Science and Technology, 3, 440

\bibitem[{Lee(2009)}]{YunShikLeeBook2009}
Lee, Y.-S. 2009, {Principles of Terahertz Science and Technology} (New York,
  NY, USA: Springer)

\bibitem[{{Loeffler} {et~al.}(2005){Loeffler}, {Baratta}, {Palumbo},
  {Strazzulla}, \& {Baragiola}}]{2005A&A...435..587L}
{Loeffler}, M.~J., {Baratta}, G.~A., {Palumbo}, M.~E., {Strazzulla}, G., \&
  {Baragiola}, R.~A. 2005, \aap, 435, 587

\bibitem[{Martin(1967)}]{PhysRev.161.1.143-155.1967}
Martin, P.~C. 1967, Physical Review, 161, 143

\bibitem[{{Mastrapa} {et~al.}(2009){Mastrapa}, {Sandford}, {Roush},
  {Cruikshank}, \& {Dalle Ore}}]{2009ApJ...701.1347M}
{Mastrapa}, R.~M., {Sandford}, S.~A., {Roush}, T.~L., {Cruikshank}, D.~P., \&
  {Dalle Ore}, C.~M. 2009, \apj, 701, 1347

\bibitem[{McGuire {et~al.}(2016)McGuire, Ioppolo, Allodi, \&
  Blake}]{PhysChemChemPhys.18.30.20199.2016}
McGuire, B.~A., Ioppolo, S., Allodi, M.~A., \& Blake, G.~A. 2016, Physical
  Chemistry Chemical Physics, 18, 20199

\bibitem[{Nyquist(1928)}]{TransAmerInstElectrEng.47.2.617.1928}
Nyquist, H. 1928, Transactions of the American Institute of Electrical
  Engineers, 47, 617

\bibitem[{{Ossenkopf} \& {Henning}(1994)}]{1994A&A...291..943O}
{Ossenkopf}, V. \& {Henning}, T. 1994, \aap, 291, 943

\bibitem[{{Ossenkopf} {et~al.}(1992){Ossenkopf}, {Henning}, \&
  {Mathis}}]{1992A&A...261..567O}
{Ossenkopf}, V., {Henning}, T., \& {Mathis}, J.~S. 1992, \aap, 261, 567

\bibitem[{{Palumbo} {et~al.}(2006){Palumbo}, {Baratta}, {Collings}, \&
  {McCoustra}}]{2006PCCP....8..279P}
{Palumbo}, M.~E., {Baratta}, G.~A., {Collings}, M.~P., \& {McCoustra}, M.~R.~S.
  2006, Physical Chemistry Chemical Physics (Incorporating Faraday
  Transactions), 8, 279

\bibitem[{{Pontoppidan} {et~al.}(2003){Pontoppidan}, {Fraser}, {Dartois},
  {Thi}, {van Dishoeck}, {Boogert}, {d'Hendecourt}, {Tielens}, \&
  {Bisschop}}]{2003A&A...408..981P}
{Pontoppidan}, K.~M., {Fraser}, H.~J., {Dartois}, E., {et~al.} 2003, \aap, 408,
  981

\bibitem[{{Preibisch} {et~al.}(1993){Preibisch}, {Ossenkopf}, {Yorke}, \&
  {Henning}}]{1993A&A...279..577P}
{Preibisch}, T., {Ossenkopf}, V., {Yorke}, H.~W., \& {Henning}, T. 1993, \aap,
  279, 577

\bibitem[{Preu {et~al.}(2011)Preu, D{\"o}hler, Malzer, Wang, \&
  Gossard}]{CW_THz_REV_2011}
Preu, S., D{\"o}hler, G.~H., Malzer, S., Wang, L.~J., \& Gossard, A.~C. 2011,
  Journal of Applied Physics, 109, 061301

\bibitem[{Pupeza {et~al.}(2007)Pupeza, Wilk, \& Koch}]{OptExp.15.7.4335.2007}
Pupeza, I., Wilk, R., \& Koch, M. 2007, Optics Express, 15, 4335

\bibitem[{{Ron} \& {Schnepp}(1967)}]{1967JChPh..46.3991R}
{Ron}, A. \& {Schnepp}, O. 1967, \jcp, 46, 3991

\bibitem[{Tukey {et~al.}(1986)Tukey, Cleveland, \& Brillinger}]{TukeyBook.1986}
Tukey, J., Cleveland, W.~S., \& Brillinger, D.~R. 1986, {The Collected Works of
  John W. Tukey. Volume I: Time Series, 1949-1964 (Wadsworth
  Statistics/Probability Series) 1st Edition} (Wadsworth Advanced Books \&
  Software)

\bibitem[{{Urso} {et~al.}(2016){Urso}, {Scir{\`e}}, {Baratta}, {Compagnini}, \&
  {Palumbo}}]{2016A&A...594A..80U}
{Urso}, R.~G., {Scir{\`e}}, C., {Baratta}, G.~A., {Compagnini}, G., \&
  {Palumbo}, M.~E. 2016, \aap, 594, A80

\bibitem[{{Van Exter} {et~al.}(1989){Van Exter}, Fattinger, \&
  Grischkowsky}]{Grischkowsky_APL_1989}
{Van Exter}, M., Fattinger, C., \& Grischkowsky, D. 1989, Applied Physics
  Letters, 55, 337

\bibitem[{{Warren} \& {Brandt}(2008)}]{2008JGRD..11314220W}
{Warren}, S.~G. \& {Brandt}, R.~E. 2008, Journal of Geophysical Research
  (Atmospheres), 113, D14220

\bibitem[{Zaytsev {et~al.}(2015)Zaytsev, Gavdush, Chernomyrdin, \&
  Yurchenko}]{IEEETransTerSciTech.5.5.817.2015}
Zaytsev, K., Gavdush, A., Chernomyrdin, N., \& Yurchenko, S. 2015, IEEE
  Transactions on Terahertz Science and Technology, 5, 817

\bibitem[{Zaytsev {et~al.}(2014)Zaytsev, Gavdush, Karasik, Alekhnovich, Nosov,
  Lazarev, Reshetov, \& Yurchenko}]{JApplPhys.115.19.193105.2014}
Zaytsev, K., Gavdush, A., Karasik, V., {et~al.} 2014, Journal of Applied
  Physics, 115, 193105

\end{thebibliography}

\end{document}